\documentclass[acmsmall,dvipsnames,svgnames,review=false,preprint]{acmart-arxiv}

\usepackage{booktabs} \usepackage{pgfplots}
\pgfplotsset{compat=newest}
\usepackage{xcolor}
\usepackage{siunitx}
\usepackage{multirow}
\usepackage{xspace}
\usepackage{balance}
\usepackage[font=rm]{caption}
\usepackage{algorithm}
\usepackage[noend]{algpseudocode}
\usepackage{verbatim}
\usepackage{graphicx}
\usepackage{enumitem}
\usepackage{shortvrb}
\usepackage{tikz}
\usepackage{subfig}
\usepackage{tabularx}
\usepackage{soul}

\usetikzlibrary{matrix, decorations.pathreplacing, calc, positioning, backgrounds}

\setlist[itemize,1]{leftmargin=5mm,itemsep=0mm}
\setlist[enumerate,1]{leftmargin=5mm,itemsep=0mm}

\newcommand{\etvbmw}{\textsf{ET-VBMW}}

\newcommand{\wand}{\textsf{WAND}}
\newcommand{\bmw}{\textsf{BMW}}
\newcommand{\vbmw}{\textsf{VBMW}}
\newcommand{\maxscore}{\textsf{MaxScore}}
\newcommand{\jass}{\textsf{JASS}}
\newcommand{\jassvtwo}{\textsf{JASSv2}}
\newcommand{\jassex}{\textsf{JASS-E}}
\newcommand{\jassag}{\textsf{JASS-A}}
\newcommand{\pisa}{\textsf{PISA}}

\newcommand{\greedy}{\textsf{Overshoot}}
\newcommand{\cautious}{\textsf{Undershoot}}
\newcommand{\judicious}{\textsf{Predictive}}
\newcommand{\adaptive}{\textsf{Reactive}}
\newcommand{\fixed}{\textsf{Fixed}}
\newcommand{\tmax}{\ensuremath{t_{\mbox{\scriptsize{max}}}}}

\newcommand{\bp}{\textsf{BP}}
\newcommand{\random}{\textsf{Random}}

\newcommand{\default}{\textsf{Default}}
\newcommand{\clustered}{\textsf{Clustered}}

\newcommand{\ltrr}{\textsf{LTRR}}
\newcommand{\oracle}{\textsf{Oracle}}
\newcommand{\boundsum}{\textsf{BoundSum}}
\newcommand{\bndsum}{\textsf{BndSum}}

\newcommand{\violatex}{\color{WildStrawberry}}
\newcommand{\withinx}{\color{RoyalBlue}}

\newcommand{\delete}[1]{{\color{WildStrawberry}\st{#1}}}
\newcommand{\newtext}[1]{{\color{RoyalBlue}#1}}

\newcommand{\simdbp}{\textsf{SIMD-BP128}}
\newcommand{\gegs}{\textsf{SIMD-GEG}}

\newcommand{\nextgeqd}{{\textsf{NextGEQ$(d)$}}}
\newcommand{\nextgeq}{{\textsf{NextGEQ}}}
\newcommand{\seekgeqd}{\textsf{SeekGEQ$(d)$}}
\newcommand{\seekgeq}{\textsf{SeekGEQ}}

\newcommand{\method}[1]{{\sf{#1}}}
\newcommand{\gb}[1]{{\mbox{$#1$~GiB}}}

\newcommand{\var}[1]{\mbox{\emph{#1}}}

\def\D{\hphantom{1}}

\newcommand{\myparagraph}[1]{\vspace*{-0.25ex}\paragraph*{\hspace*{-\parindent}\normalsize\bf#1}}
\newcommand{\mycaption}[1]{\caption{{\rm{#1}}}}

\usepackage{xcolor}

\newcommand{\joel}[1]{{\color{orange}{\bf{Joel says:}} \emph{#1}}}
\newcommand{\alistair}[1]{{\color{purple}{\bf{Alistair says:}} \emph{#1}}}
\newcommand{\matthias}[1]{{\color{DarkCyan}{\bf{Matthias says:}} \emph{#1}}}
\newcommand{\todo}[1]{{\color{green}{[[#1]]}}}

\newcommand{\marker}[1]{\textcolor{NavyBlue}{\fcolorbox{NavyBlue}{Apricot}{\textbf{R-#1}}}}

\newcommand{\collection}[1]{\method{#1}}
\newcommand{\govtwo}{\collection{Gov2}}
\newcommand{\clueweb}{\collection{ClueWeb09B}}
\newcommand{\bigclueweb}{{\clueweb\,$\times\,P$}}

\newcommand{\Contrib}{{\mathcal{C}}}
\newcommand{\Query}{{\mathcal{Q}}}
\newcommand{\Sim}{{\mathcal{S}}}

\newcommand{\Termmaxr}{U_{t,i}}

\newcommand{\daat}{\sc{DaaT}}
\newcommand{\saat}{\sc{SaaT}}

\renewcommand{\delete}[1]{}
\renewcommand{\newtext}[1]{#1}
\renewcommand{\joel}[1]{}
\renewcommand{\alistair}[1]{}
\renewcommand{\matthias}[1]{}
\renewcommand{\todo}[1]{}
\renewcommand{\marker}[1]{}

\setcopyright{acmlicensed}

\acmDOI{}

\acmISBN{}

\acmJournal{TOIS}
\copyrightyear{2021}
\acmMonth{4}
\acmYear{2021}
\acmVolume{1}
\acmNumber{1}
\acmArticle{1}
\acmPrice{15.00}
\acmDOI{10.1145/nnnnnnn}

\acmISBN{}

\begin{document}
\title{Anytime Ranking on Document-Ordered Indexes}

\author{Joel Mackenzie}
\orcid{0000-0001-7992-4633} 
\affiliation{\institution{The University of Melbourne}
  \city{Melbourne} 
  \state{Australia} 
}

\author{Matthias Petri}
\orcid{0000-0002-0054-9429}
\affiliation{\institution{Amazon Alexa}
  \city{Manhattan Beach} 
  \state{CA, USA} 
}

\author{Alistair Moffat}
\orcid{0000-0002-6638-0232}
\affiliation{
  \institution{The University of Melbourne}
  \city{Melbourne}
  \state{Australia}
}

\begin{abstract}
Inverted indexes continue to be a mainstay of text search engines,
allowing efficient querying of large document collections.
While there are a number of possible organizations, document-ordered
indexes are the most common, since they are amenable to various query
types, support index updates, and allow for efficient dynamic pruning
operations.
One disadvantage with document-ordered indexes is that high-scoring
documents can be distributed across the document identifier space,
meaning that index traversal algorithms that terminate early might
put search effectiveness at risk.
The alternative is impact-ordered indexes, which primarily support
top-$k$ disjunctions, but also allow for {\emph{anytime}} query
processing, where the search can be terminated at any time, with
search quality improving as processing latency increases.
Anytime query processing can be used to effectively reduce
high-percentile tail latency which is essential for operational
scenarios in which a service level agreement (SLA) imposes response
time requirements.
In this work, we show how document-ordered indexes can be organized
such that they can be queried in an anytime fashion, enabling strict
latency control with effective early termination.
Our experiments show that processing document-ordered topical
segments selected by a simple score estimator outperforms existing
anytime algorithms, and allows query runtimes to be accurately
limited in order to comply with SLA requirements.
 \end{abstract}

\ccsdesc[500]{Information systems~Retrieval efficiency}
\ccsdesc[500]{Information systems~Search engine architectures and scalability}

\keywords{Tail latency, dynamic pruning, query processing, web search}

\maketitle

\section{Introduction}
\label{sec-introduction}

The enormous volume of web search queries processed every second of
every day makes efficient first-phase top-$k$ query processing
against web-scale indexes critically important, and means that even
small percentage reductions in computation can translate into
large monetary savings in terms of hardware and energy costs.
To harvest those savings, enhancements to query processing have been
developed over several decades, seeking to improve on the exhaustive
``process all postings for all query terms'' starting point.
In addition, to ensure that system responsiveness is maintained, it
is usual for a service level agreement (SLA) to be in place,
typically mandating that some fraction of queries (for example, 95\%)
must be responded to within some stipulated upper time limit.

One important strand of development has led to dynamic pruning
techniques such {\maxscore} {\citep{tf95-ipm}}, {\wand}
{\cite{bc+03-cikm}}, {\bmw} {\citep{ds11-sigir}}, and {\vbmw}
{\citep{mo+17-sigir}}; these mechanisms process only a subset of the
query terms' postings, but in a manner that guarantees that the
top-$k$ document listing computed is {\emph{safe}}, and true to the
scores that would emerge from exhaustive processing.

A second strand of development has pursued non-safe approaches,
including the quit/continue heuristics {\citep{mz96-tois}}; and
approximations that deliberately over-estimate the current heap entry
threshold (when maintaining the top-$k$ set) {\citep{ccm16-irj,
cclmt17wsdm, bc+03-cikm, tmo13-wsdm, mto12-sigir-poster}}.
In related work, researchers have sought bounds on the query's final
$k$\,th largest document score {\citep{dm+17-irj,kt18sigir,ya19-cikm}}, or to
provide a conservative estimate of it
{\citep{pm+19-sigir,cikm20msss}}, seeking to bypass fruitless early
work when dynamic pruning mechanisms are in play. 

The third strand of work proposes the use of impact-ordered indexes
and score-at-a-time processing
{\citep{akm01-sigir,am06-sigir,lt15-ictir,cclmt17wsdm}}
which place the ``high impact'' postings for each term at the front
of the postings list, allowing them to be processed first.
Score-at-a-time processing allows effective {\emph{anytime ranking}}
{\citep{akm01-sigir}}, an important consideration if response time
expectations are imposed via an SLA, and query processing must be
interruptible.
If sufficient computation time is allowed that all of the postings
can be processed, then best-possible effectiveness is achieved; and
if the processing must be cut short because of the SLA, a best-so-far
approximation to the final ordering is generated.
If the impact scores are quantized to integers, which is the usual
case, then score-at-a-time approaches are non-safe, with the
quantization level setting the fidelity.

Selective search {\citep{ak15-tois,kccm16-irj}} provides a fourth
strand of ideas.
In a selective search system, the set of documents is partitioned
into topical shards at index construction time, with each shard
ideally containing a set of topically-related documents.
Each incoming query is considered by a broker process, which predicts
which subset of the
shards should be consulted for that query.
The query is then processed against the selected shards, and a result
synthesized from the shards' result sets.
Within each shard the search can be carried out using any of the
techniques already described, including, for example, {\wand} pruning
{\citep{kc+16-ecir}}.
The fact that only some of the shards process each query means that
selective search cannot be safe, but that it is efficient in terms of
workload.

Independent of these four strands are two further considerations,
based on {\emph{data volume}}, and on {\emph{query volume}}.
If more data must be handled than can be managed by a single machine,
then {\emph{partitioning}} across multiple machines is required, with
all of them processing each query, and their answer sets merged into
a single result set.
Likewise, if the query load is larger than can be handled on a single
machine (or cluster of machines, if the collection has been
partitioned), then the collection must be {\emph{replicated}} as many
times as necessary, and further machines (or clusters of machines)
brought into service.
Index and document compression techniques can be employed to increase
the volume of data on each machine or cluster, whereas efficient
query processing approaches of the type summarized above reduce the
number of replicates required.
Hence, both types of enhancement save hardware and energy costs.
Note also that partitioning alone (for example, by random document
assignment) does not improve throughput -- it is a coping technique
to deal with large data volumes; and while it has the attractive side
effect of reducing query latency, it does not reduce the overall
computation footprint.
Indeed, when pruning techniques are being used, random partitioning
increases the total work undertaken per query, and hence reduces the
query throughput that is possible within a given overall hardware
envelope.
That is, regardless of how large the document collection, or how
great the query rate to be supported, the computations on each
machine, supporting one machine's worth of data and one machine's
worth of the query load, should be designed to be as efficient as
possible.

\subsection{Contribution}

In that context, our purpose in this paper is to examine one
processing node in a (possibly) partitioned and (possibly) replicated
system, and revisit the question as to what form of index and what
mode of processing allows query handling with the least computational
cost.
Our proposal contains elements drawn from several of the four strands
of development listed above, and employs a topically-segmented
document-reordered index, and a modified form of document-at-a-time
processing.
{\newtext{As a very brief summary, we propose that the documents
managed at any processing node be reordered into topically coherent
ranges; that each of those ranges have additional index information
associated with it that can be used to determine a query-dependent
``likely usefulness'' score; and that the query then be resolved by
processing the ranges in decreasing order of the query-dependent
score, thereby enhancing the savings arising from dynamic pruning.}}
{\marker{3.6}}
We also take as axiomatic the requirement for an enforceable SLA on
response time, and hence seek an anytime processing approach,
provided that the ``any'' is a reasonable allowance{\newtext{, and
further refine our proposed mechanism to support early termination in
a manner that is sensitive to the stipulated SLA}}.
{\marker{3.6}}
In particular, we show that:
\begin{itemize}
\item
	a sequential (rather than parallel) implementation of
	selective search can be supported by an augmented
	document-ordered index based on a {\emph{cluster skipping
	inverted index}} {\cite{ad+08-tois, hka17ecir}} with cluster
	selection possible via a low-cost heuristic; 

\item
	secondary document reordering via recursive graph bisection
	{\citep{dk+16-kdd}} results in localized spans of document
	numbers in which the top-$k$ heap thresholds grow quickly,
	accelerating dynamic pruning mechanisms; and

\item
	a further low-cost heuristic allows an execution-time SLA to
	be complied with, making effective anytime ranking
	feasible.

\end{itemize}
We also report a detailed experimental evaluation of our proposal,
comparing it to a range of baseline approaches including
score-at-a-time processing on an impact-ordered index, and show that:
\begin{itemize}
\item
	document reordering also dramatically accelerates
	score-at-a-time query processing, thereby providing a
	highly-competitive reference system not previously noted in
	the literature; and

\item
	in terms of ``best-so-far'' processing, measured by computing
	retrieval effectiveness achieved across a range of SLA
	options, our new anytime processing mode on document-ordered
	indexes outperforms score-at-a-time processing over
	the majority of the
	likely response-time range.

\end{itemize}
The modified index structure can still support traditional
querying modes, such as efficient Boolean conjunction.

\section{Background}
\label{sec-background}

Search systems are typically implemented via multi-stage ranking
cascades, as a balance between efficiency requirements and the desire
to generate effective document rankings {\cite{wlm11sigir,
cgbc17-sigir, wds16-sigir}}.
In the first phase the goal is to identify a set of possibly relevant
documents using a simple ranking function.
Those candidates are then re-ranked as required by more costly
subsequent phases.
First-stage rankers typically assume term independence, using
bag-of-words models to find a top-$k$ document list.
Given a query $\Query = \{t_1, t_2, \dots, t_n\}$, each document $d$
is scored as a sum of contributions over terms:
\[
\Sim(\Query, d) = \sum^n_{i=1}{\Contrib(t_i, d)} \, ,
\]
where $\Contrib(t_i, d)$ is the contribution to $d$ from $t_i$.
This formulation captures many common bag-of-words ranking models, 
including {\sf{BM25}} {\cite{rz09fntir}}, language models {\cite{pc98-sigir}}, 
and approaches based on divergence from randomness {\citep{ar02-tois}}.

\subsection{Indexing and Query Processing}
In order to maintain rapid response times to incoming queries, search engines
rely on efficient indexing and query processing techniques. We now briefly
review some of the key techniques for facilitating efficient and scalable
query processing.

\myparagraph{Document-Ordered Indexes}

An inverted index contains a {\emph{postings list}} for each unique
term $t$ that appears, containing the document numbers $d$ containing
$t$.
Scoring then involves traversing each of the query terms' postings
lists, adding together each document's contributions.
In a {\emph{document-ordered index}} the postings in each list are
stored in increasing order of document identifier.
Typically, a document-ordered index will store document identifiers
and term frequency ({\emph{tf}}) information separately, so that
document identifiers can be decompressed {\emph{without}} requiring
that the {\emph{tf}} data be decompressed as well.
Postings lists are usually organized into fixed-length blocks that
are compressed independently.

An important optimization for document-ordered indexes involves
reassigning the document identifiers such that similar documents
appear close together in the document identifier space, thereby
reducing the size of the index and increasing query throughput
{\cite{mss19-ecir}}.
While many reordering techniques are possible {\cite{mm+19-ecir}},
the most common heuristic involves ordering web documents by URL
{\citep{s07-ecir}}.
More recently, {\citet{dk+16-kdd}} describe a recursive graph
partitioning approach which represents the current state-of-the-art
for document identifier reassignment on both graphs and inverted
indexes {\cite{dk+16-kdd, mm+19-ecir, mpm21-sigir}}.

\myparagraph{Document-at-a-Time Traversal}

\begin{figure}[t]
\centering
\includegraphics[width=0.70\textwidth]{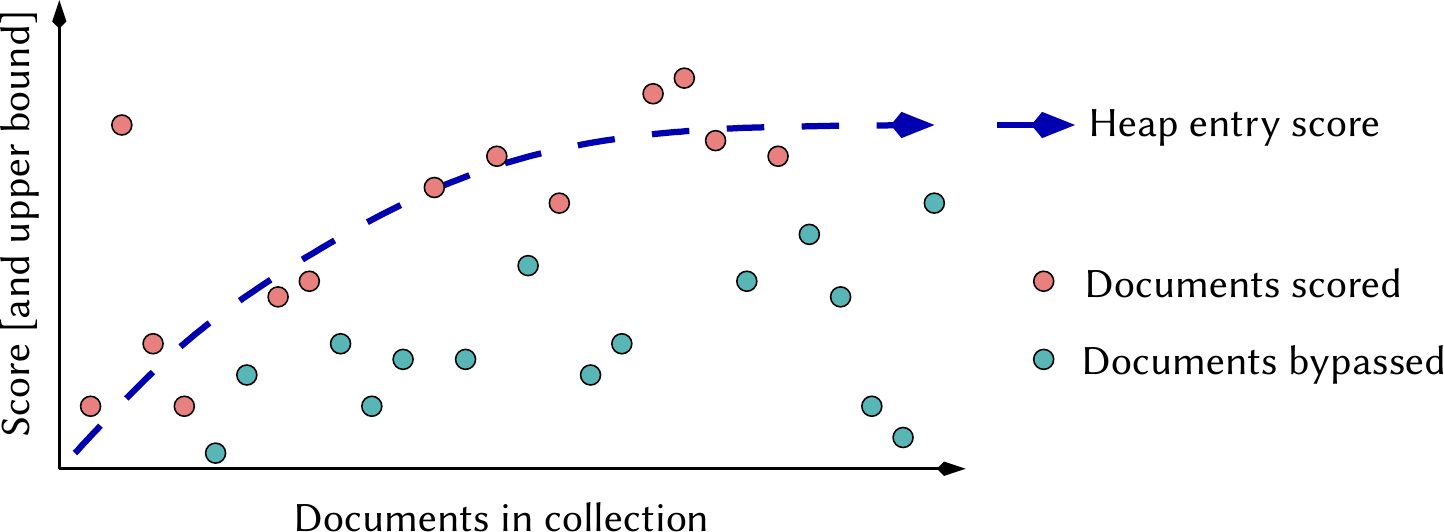}
  \\[1.5ex]
(a) Normal pruning, with the heap threshold in blue.
  \\[3ex]
\includegraphics[width=0.70\textwidth]{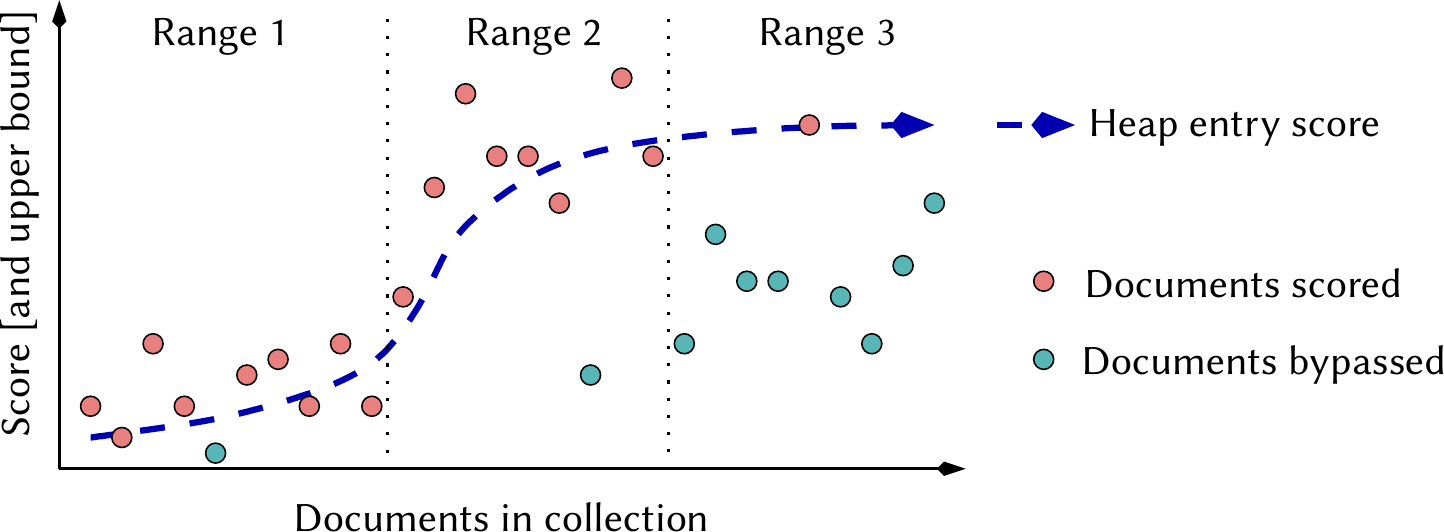}
  \\[1.5ex]
(b) Pruning in a topically-coherent collection.
  \\[3ex]
\includegraphics[width=0.70\textwidth]{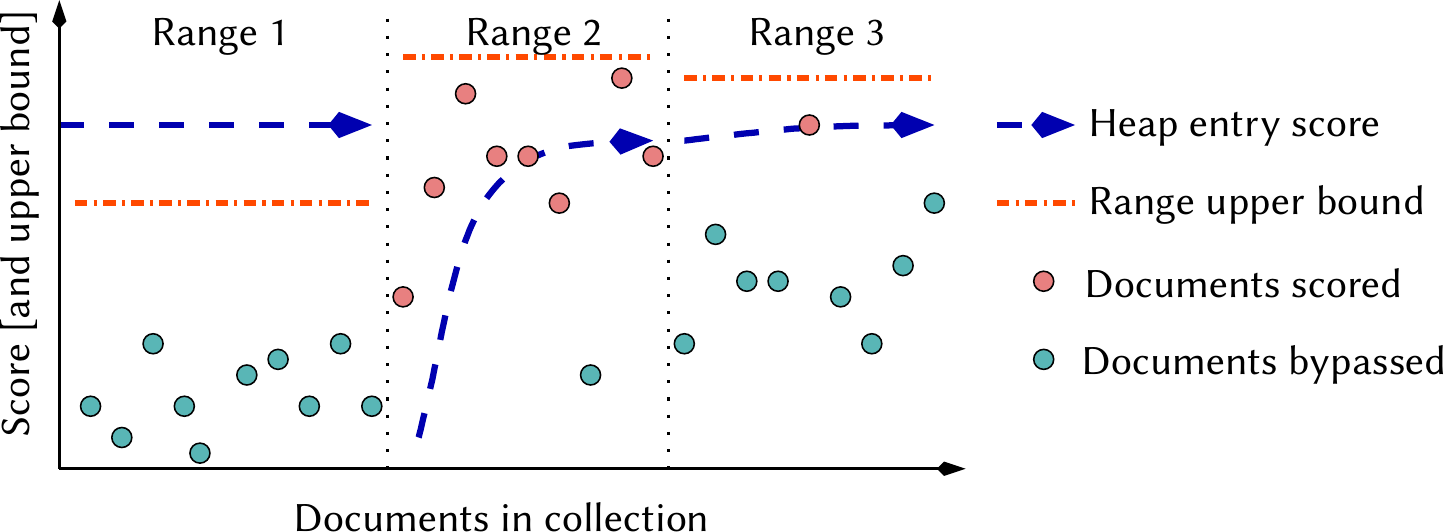}
  \\[1.5ex]
(c) Range-prioritized processing, based on upper bounds.
\\
\mycaption{Dynamic pruning: (a) on an unordered document collection,
with high-scoring documents scattered throughout; (b) on a document
collection in which the documents are topically-clustered into three
coherent ranges; and (c) on the same document collection, but with
range selection applied.
In parts (b) and (c) it is assumed the documents of interest were
primarily located in range~2.
{\newtext{Those two parts of the figure are discussed in
Section~\ref{sec-concept}.}}
\label{fig-example}}
\end{figure}

A document-ordered index is typically processed in a
{\emph{document-at-a-time}} ({\daat}) manner, with all of the
postings for each document processed at the same time.
A min-heap is used to record the top-$k$ documents encountered so
far, with a threshold value $\theta$ tracking the lowest score among
those $k$.
Within this framework a range of {\emph{dynamic pruning}} algorithms
can be applied, with the aim of bypassing a large fraction of the
postings.
These approaches include {\maxscore} {\cite{tf95-ipm}} and {\wand}
{\cite{bc+03-cikm}}, both of which pre-compute the largest value of
$\Contrib(t, d)$ for each term $t$, and store it as $U_t$.
Since each document score is computed as a sum of contributions
over query terms, the $U_t$ values can be used to generate a loose
upper-bound score for each document; and documents which cannot
achieve a score greater than $\theta$ can be bypassed.
Subsequent enhancements to the {\wand} mechanism include block-max
{\wand} ({\bmw}) {\citep{ds11-sigir}}, which stores an upper bound
per fixed-length block of postings, and thus provides localized score
bounds that are more likely to be accurate; and variable block-max
{\wand} ({\vbmw}) {\citep{mo+17-sigir}} in which the block lengths
are adaptive in response to the local distribution of score
contributions in that section of the postings list.
Figure~\ref{fig-example}(a) illustrates the way in which the value of
$\theta$ increases as documents are processed, and how (all other
things being equal) an increasing fraction of the documents are
bypassed by dynamic pruning approaches as each query is processed.
{\newtext{(Figures~\ref{fig-example}(b) and~\ref{fig-example}(c) will
be discussed in Section~\ref{sec-concept}.)}}
{\marker{3.10}}
{\citet{pcm13-adcs}} and {\citet{mm+18-ecir}} provide examples that show the
way in which $\theta$ grows on actual queries.

Pointer advance in a document-ordered postings list is achieved using
a {\nextgeqd} operator, which forwards the list's cursor from its
current position to that of document $d$, or the next posting after
it if $d$ is not present.
Efficient implementations of {\nextgeq} are crucial to efficient
query processing, and are why document-ordered indexes store postings
lists as a series of independent blocks.
For each block, the maximum document identifier $d$ is stored in a
secondary structure to allow efficient skipping to blocks containing
a specific $d$.
While {\nextgeq} is used to skip {\emph{forwards}} from the current
position, the underlying skip information stored along block
boundaries allows bidirectional seeking to {\emph{any}} arbitrary
block efficiently~{\cite{ov14-sigir}}.

\myparagraph{Impact-Ordered Indexes}

In an {\emph{impact-ordered index}} the postings are ordered by
decreasing impact, where the impact of $t$ in $d$ is an
integer-quantized approximation of ${\Contrib(t, d)}$
{\citep{akm01-sigir,am06-sigir}}.
Each segment of each postings list contains an integer impact
followed by an ordered list of document identifiers that all share
that impact.
The impact scores are computed at the time the index is constructed,
and are a form of pre-computation, with a global mapping used to
approximate each real-valued ${\Contrib(t, d)}$ by a $b$-bit integer
in the range $[1, 2^{b-1}]$.
That is, each postings list consists of up to $2^{b-1}$ segments,
each containing a sequence of sorted document identifiers.

The use of quantized approximations means that this type of index
might yield different documents scores to when the same arithmetic is
carried out using floating-point arithmetic, but that difference can
be controlled via the choice of $b$, and experiments have shown that
values in the range $8$--$10$ are sufficient for large collections
{\citep{akm01-sigir,am06-sigir, ct+13-cikm, cclmt17wsdm}}.

\myparagraph{Score-at-a-Time Traversal}

Impact-ordered indexes are processed using a score-at-a-time
({\saat}) traversal {\citep{akm01-sigir,am06-sigir}}.
Segments are processed from the set of postings lists in strictly
non-increasing impact order, regardless of term, so that the largest
contributions for any particular document $d$ are accumulated first,
in a ``best foot forward'' approach.
Processing each segment involves decompressing the set of document
identifiers, and adding the current impact score to each
corresponding document accumulator.
Once all of the segments have been processed, the accumulator table
contains the scores for all documents, and the top-$k$ set can be
identified via a heap or other suitable priority queue. 
A range of early termination heuristics are available for improving the 
efficiency of {\saat} traversal, most of which rely on setting an upper limit
on the number of postings to process {\cite{msc17-adcs,am06-sigir,lt15-ictir}}.

\myparagraph{Anytime Ranking}

Impact-ordered indexes and {\saat} retrieval combine to allow
{\emph{anytime processing}} {\citep{akm01-sigir,lt15-ictir}}, a mode
in which each query is given a fixed quantum of computation time, and
within that time limit must respond with a ``best endeavor'' document
ranking.
This is achieved by establishing an empirical relationship between
postings scored (measured in millions, perhaps) and time taken
(milliseconds, perhaps), and upon receipt of each query, determining
a set of highest-impact segments to be processed that fit within the
estimated postings limit.
The ranked result that is returned will not be safe, of course.
{\newtext{But it will be more useful than what could be achieved if the same
limit on postings was enforced in {\daat} mode. This is because {\daat}
traversal has no notion of ``priority'', simply scanning through the
postings lists from start to finish, and hence would be unable to supply any 
answers at all from the unprocessed tail of the index.}} {\marker{3.9}}
Indeed, recent work has examined the use of impact-ordered indexes
and {\saat} retrieval in a hybrid system with document-ordered
indexes and {\daat} retrieval for reducing tail latency
{\cite{mcbccl19wsdm}}.
One potential weakness with this processing mode arises for long
queries or queries with many long postings lists, as only a small
percentage of the total impacts may be processed within the given
budget {\cite{msc17-adcs}}.

\subsection{Topical Shards and Selective Search}

Where a collection is too large to fit on one machine, it must be
partitioned into {\emph{shards}}, as noted in
Section~\ref{sec-introduction}.
Partitioning based on random assignment of documents has
load-balancing and response-time latency advantages.
But partitioning by topic -- referred to as {\emph{selective search}}
{\citep{ak15-tois,ah+13-sigir,kccm16-irj}} -- allows greater
throughput to be achieved, provided that a suitable subset of the
shards can be identified as providing a sufficient response to each
query {\cite{kt+12-cikm,dkc17-sigir}}.
There are many components to be managed in a selective search system,
including assignment of documents to shards in some cohesive manner
and building features for shard selection at indexing time; and then,
for each query, determining a rank-ordering of shards for that query
via the stored features, and deciding how far down that ranking to
proceed {\cite{mx+18-sigir}}.
Once the selected subset of shards has executed the query in parallel
and generated their top-$k$ answer lists, the
{\newtext{coordinating}} broker {\newtext{process}} merges those
answers into an overall top-$k$ answer.
Within each shard dynamic pruning algorithms can be employed, and the
computational gains from topical sharding and pruning are additive
{\cite{kc+16-ecir}}.
Similar work has considered processing index shards on a
{\emph{site-by-site}} basis, where only a subset of selected websites
are examined {\cite{ad+08-sigir}}.

\myparagraph{Cluster-Based Indexes}

Cluster-based retrieval precedes selective search, with systems
proposed as early as the 1970s.
In a cluster-based index each postings list is arranged into a set of
related ``clusters'', housing sets of related documents, and
facilitating localized selective search {\cite{cad04-is}}.
To traverse such a {\emph{cluster skipping inverted index}}, each
query is first examined to determine which clusters should be
searched, and then those portions of the postings lists are used to
select documents.
{\citet{ad+08-tois}} examine the efficiency of cluster-based search
systems with {\emph{term-at-a-time}} retrieval, including exploration
of a cluster-based document identifier reassignment technique, which
assigns identifiers to documents within each cluster consecutively,
and orders the clusters according to their creation order.
A similar idea was recently investigated by {\citet{srs19-ecir}}, who
reorder documents by {\emph{hit count}}, using a query log to
determine the documents that appear most regularly in the top-$k$
lists.
A resource allocator task estimates the clusters to search, and
how deep to search within each cluster, by providing a global
ordering cutoff.

A similar approach is discussed by {\citet{hka17ecir}}, who improve
the efficiency of selective search via better load balancing.
Instead of distributing all topically related documents to a single
server (as is typical in selective search), they instead distribute a
fraction of {\emph{each}} topical shard to {\emph{every}} server.
When each query arrives it is forwarded to all servers, but with only
selected chunks of each index searched, rather than the full index.

\subsection{Service Level Agreements}

In commercial settings it is common practice to precisely specify
requirements within which a service must be provided via a
contractual commitment, referred to as a {\emph{service level
agreement}} (SLA).
An SLA stipulates the guarantees provided by a service to its clients
with respect to performance metrics such as latency or durability.
Response-time SLA metrics are generally codified in terms of tail
latency at specific percentiles, as tail latency is more
representative of the worst case response latency than median
latency.
For example, Azure Cosmos
DB\footnote{\url{https://azure.microsoft.com/en-us/support/legal/sla/cosmos-db/v1_3/},
retrieved October 2020} provides for a tail 99\,th percentile
($P_{99}$) read latency SLA of 10~milliseconds.

Similarly, search engines often operate under strict tail latency
constraints, with the time of each phase of the retrieval cascade
bounded so as to ensure an SLA for the system as a whole
{\cite{rc+13-wsdm, hkhec16-tweb}}, and to improve the predictability
of the end-to-end search process.
In addition, complex neural models are often used in components such
as query expansion, query understanding, and document re-ranking
{\cite{gcbc18-wsdm, kz20-sigir, mn+20-sigir}}, further restricting
the time available to the first stage of the retrieval pipeline.

A number of authors have examined SLAs at various levels of the
distributed search architecture.
For example, {\citet{yher15-sigir}} investigated aggregation policies
to meet SLAs at the level of the {\emph{results aggregator}}, where
top-$k$ rankings from a number of {\emph{index server nodes}} (ISNs)
are combined before being presented to users.
This idea is also discussed by {\citet{db13-cacm}}, who note that
users may have a better search experience if they are given slightly
incomplete results, but with improved response time.
Consideration has also been given to improving tail latency
{\emph{within}} each ISN.
One area of focus has been on latency prediction and selective
parallelization, where (predicted) long-running queries are
accelerated by employing additional worker threads {\cite{sk15-wsdm,
jk+14-sigir, hkhec16-tweb, msc17-adcs}}.
Indeed, reducing the latency of each ISN is a key requirement for
handling larger workloads, as ISNs make up over 90\% of the total
hardware resources for large-scale search deployments, and can
account for more than two-thirds of end-to-end query latency
{\cite{jk+14-sigir}}.
 
\section{Anytime Ranking Techniques}
\label{sec-concept}

This section describes how a document-ordered index can be
reorganized in a way that still allows normal {\daat} processing to
be carried out, as well as enabling a new {\emph{anytime
document-at-a-time}} processing mode.
{\newtext{The ordering of the documents within each ISN plays a key
part, and is discussed first; then we consider how dynamic pruning
approaches such as {\wand} might be affected by topical clustering of
documents, leading to the introduction of the key innovation, that of
{\emph{range-aware processing}}.
Achieving range-aware processing requires additional information to
be retained in the index, and necessitates a range selection
heuristic so that the ranges can be processed in a suitable order.
Both of those facets are also addressed in this section, along with a
first approach to anytime ranking.
Experimental measurement of the new approach is provided in
Section~\ref{sec-setup} and Section~\ref{sec-experiments}; and
Section~\ref{sec-anytime} then considers anytime ranking in detail,
including further experiments in which latency limits are
increasingly constrained.}}
{\marker{3.11}} 

\myparagraph{Document Arrangement}

\begin{figure}[t]
  \centering
  \includegraphics[width=0.95\columnwidth]{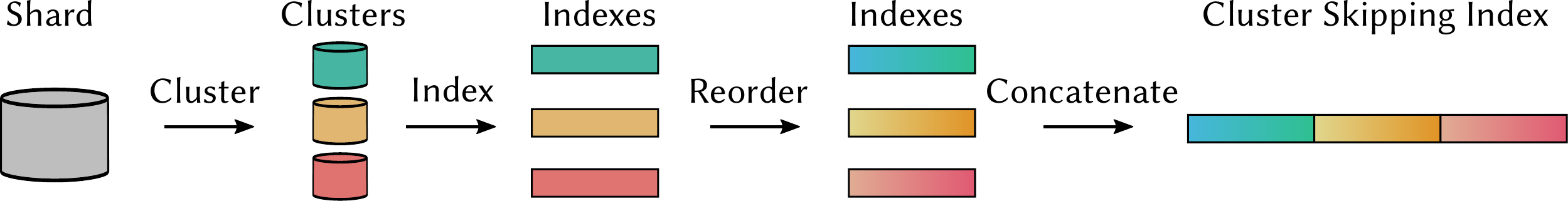}
  \mycaption{An overview of the process for building the cluster
  skipping inverted index at a given index server node.
  The shard of the global corpus being handled at this single ISN is
  segmented into a number of topical clusters, each of which is then
  indexed individually.
  Next, each cluster index is reordered internally, to improve its
  internal coherence; and finally, the cluster indexes are
  concatenated together to form the ISN index, structured as a
  sequence of document ranges, with each range corresponding to one
  of the clusters.
  The same process is repeated at each ISN.
  \label{fig-arrangement}}
\end{figure}

The first {\newtext{step toward our proposed
processing methodology}}{\marker{3.11}} is that a suitable document
ordering be created, such that documents that are ``like'' each other
in terms of content (and hence in terms of numeric similarity score
to any particular query) tend to also be ``near'' each other in the
document space.
The technique we exploit in our experiments is a composition of two
approaches: the topical clustering mechanism of {\citet{ak15-tois}}
and {\citet{kccm16-irj}}; together with the recursive graph
partitioning technique of {\citet{dk+16-kdd}} applied to each of the
clusters once they have been formed.
The whole-of-collection reordering is then generated by concatenating
the internally-reordered clusters together in any sequence.
That is, the final document ordering is created by first forming
topical clusters, then reordering within each topical cluster, and
finally bringing them back together.
Figure~\ref{fig-arrangement} gives an overview of this process.
To avoid ambiguity and ensure that there is no suggestion of
partition-based parallelism, we refer to the final collection as
consisting of a concatenation of document {\emph{ranges}}, with the
documents in each range being (hopefully) topically coherent, and
refer to this arrangement as a cluster skipping (or {\clustered})
inverted index.
We also explored utilizing only recursive graph partitioning without
topical clustering, and found that the addition of clustering
substantially improved range coherence.
We will explore recursive partitioning for document clustering more
carefully in future work.

\myparagraph{Range-Oblivious Processing}

Figure~\ref{fig-example}(b) illustrates what might occur as a result
of such a reordering, supposing that three document ranges (derived
from three topical clusters) are formed from the collection of
Figure~\ref{fig-example}(a).
In this example it is assumed that range~2 contains a high proportion
of the documents that score well for this query, meaning that the
heap threshold $\theta$ climbs most steeply within that range.
Other queries might be focused on the documents in range~1, or on the
documents in range~3.

As is shown in Figure~\ref{fig-example}(b), if the high-scoring
documents appear in any range other than the first one, normal
{\daat} processing through the reordered collection has the potential
to be wasteful, with the low-scoring documents in the first range
serving as a distraction to the dynamic pruning process.
There is little to be gained by reordering the collection if the
processing is {\emph{range oblivious}} in the way that standard
implementations of dynamic pruning algorithms such as {\wand},
{\maxscore}, and {\bmw} are.

\myparagraph{Range-Aware Processing}

{\makebox[0mm]{}}{\newtext{Comparing}} Figure~\ref{fig-example}(c)
{\newtext{with Figures~\ref{fig-example}(a)
and~\ref{fig-example}(c)}} motivates our proposal.
{\marker{3.10}} If it was possible to identify range~2 as being the
most ``answer rich'' range and process it first, then the
rapidly-climbing heap threshold $\theta$ would mean that fewer
documents would need to be scored from the other two ranges, and a
greater fraction of their documents could be bypassed.
More generally, if the ranges could be processed {\newtext{such that the
heap threshold increases at the fastest rate,}} the benefit
arising from dynamic pruning should be maximized.
Scoring document ranges in terms of their propensity to contain
answers is, of course, exactly what occurs in selective search.
The difference here is that the processing of the ranges is
sequential (rather than parallel), which introduces further
opportunities for computational savings.

The remainder of this section provides details of the various
components required to make this proposal a reality, discussing the
additional index information required to support access to the
document ranges; approaches for constructing query-influenced range
orderings, to maximize the benefits of dynamic pruning; and
additional techniques for terminating the processing in both a
score-safe arrangement and via a non-safe anytime heuristic.

\myparagraph{Index Organization}

\begin{figure}[t]
  \centering
  \includegraphics[width=0.55\columnwidth]{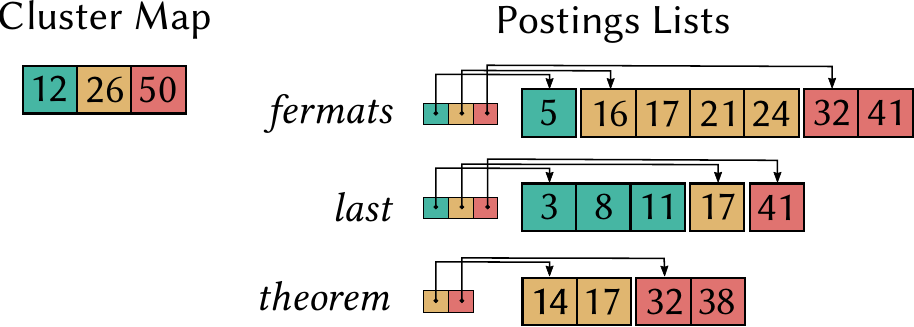}
  \mycaption{Postings list arrangement for three possible query
  terms, {\emph{``fermats''}}, {\emph{``last''}}, and {\emph{``theorem''}}.
  In addition to the postings lists, a global cluster map stores the
  last document identifier in each range. Note that the postings list range ``pointers'' are not required in
  practice (figure adapted from~\citet{hka17ecir}).
  \label{fig-architecture}}
\end{figure}

Figure~\ref{fig-architecture} shows an example of a clustered index,
with the ranges indicated by the colored shades.
A {\emph{cluster map}} vector $C = \langle c_1, c_2, ..., c_r\rangle$
records the final document number in each of the $r$ ranges, with the
$i$\,th range spanning the documents numbered $[c_{i-1}+1, c_i]$, and
with $c_0$ taken to be zero.
{\marker{3.1}} {\newtext{Our proposed processing
methodology also requires}} random access into the postings lists
{\newtext{in both the ``forwards'' (as is usual
for dynamic pruning) and ``backwards'' directions.
To that end,}} we {\newtext{introduce}} a modified
{\nextgeqd} operator {\newtext{that we denote}}
{\seekgeqd}.
The {\newtext{new}} {\seekgeqd} operator {\newtext{facilitates the locating of 
the posting for}} any
{\newtext{document}}~$d${\newtext{, regardless of the current
processing position in the postings list}}.
One {\newtext{straightforward}} implementation is to
reset the starting point to document~$0$ and then perform a
{\newtext{conventional}} {\nextgeqd}, but other implementations that
exploit locality are also possible.
All {\newtext{of the}} auxiliary information required to perform
{\seekgeqd} is already stored in order to support document bypass as
part of dynamic pruning{\newtext{, and no additional information is
required}}.
{\newtext{Then, to carry out a key step in our proposal,
namely, processing the ranges in an order that is query dependent, it
is sufficient}} access the first document of the $i$\,th range in any
of the postings lists, {\newtext{by calling}}
{\seekgeq}$(c_{i-1}+1)$.
Indeed, the sets of per-list pointers shown {\newtext{by the arrows}}
in Figure~\ref{fig-architecture} are implicit, and do not require
explicit storage {\cite{hka17ecir, cad04-is}}.

\myparagraph{Range Selection}

Range selection in our proposal serves the same role as shard
selection in selective search, and {\newtext{hence}} could be carried
out using the same strategies.
The approaches used in selective search often make use of a central
shard index (CSI), a sampling of a small percentage of documents from
each of the shards, employed as an indicator as to which shards might
contain high-scoring documents and should thus be handed the query
{\citep{ak15-tois}}.
Alternatives include combining a small number of features stored on a
per-term per-shard basis {\citep{ah+13-sigir}}, or using learned
models over a dozen or more per-term per-shard features
{\citep{dkc17-sigir, srs19-ecir, mx+18-sigir}}.
In selective search, it is necessary for both a shard ordering
and a numeric shard count to generated at the time the broker is
routing each query {\cite{mx+18-sigir, kt+12-cikm, ak15-spire}}.

In our arrangement, the fact that the processing is sequential rather
than parallel provides useful flexibility.
In particular, there is no broker, and the ``how many ranges are to
be processed'' question can be revisited after each range has been
processed, rather than decided up-front; with processing all ranges
always available as a fallback option.
That flexibility means that it is less critical that the range
ordering computed be a high-quality one, since an estimation of
quality can be monitored while the query is being evaluated.

Our proposal, which we denote {\boundsum}, is to extend whichever
dynamic pruning protocol is in play to the document ranges.
Inspired by the notion of storing term frequencies on a per-range
basis {\cite{ad+08-tois}}, a {\emph{range score bound}}, $\Termmaxr$,
is associated with each term $t$ that appears in each range $i$, and
maintained in an auxiliary data structure that is indexed by~$t$.
Each $\Termmaxr$ value is the largest score contribution arising from
$t$ for any document in the $i$\,th range, and is computed at the
time the index is constructed.
Upon receipt of a query, the sets of $\Termmaxr$ values for each term
$t$ are fetched, each a vector of $r$ values, one for each of the $r$
ranges; are added together as vectors; and then the $r$ values in the
sum are sorted into decreasing order.
Compared to executing the query against a CSI, or evaluating a
learned model using dozens of features per-term per-range, our
{\boundsum} approach based on sums of $r$-element numeric vectors is
very fast; and, as was noted above, some leeway can be permitted in
the quality of the range ordering that is generated {\newtext{-- a
bad ordering might slow execution, but need not affect retrieval
effectiveness}}.
We explore this idea further in Section~\ref{sec-experiments}.

\begin{figure*}[t]
  \includegraphics[width=0.98\textwidth]{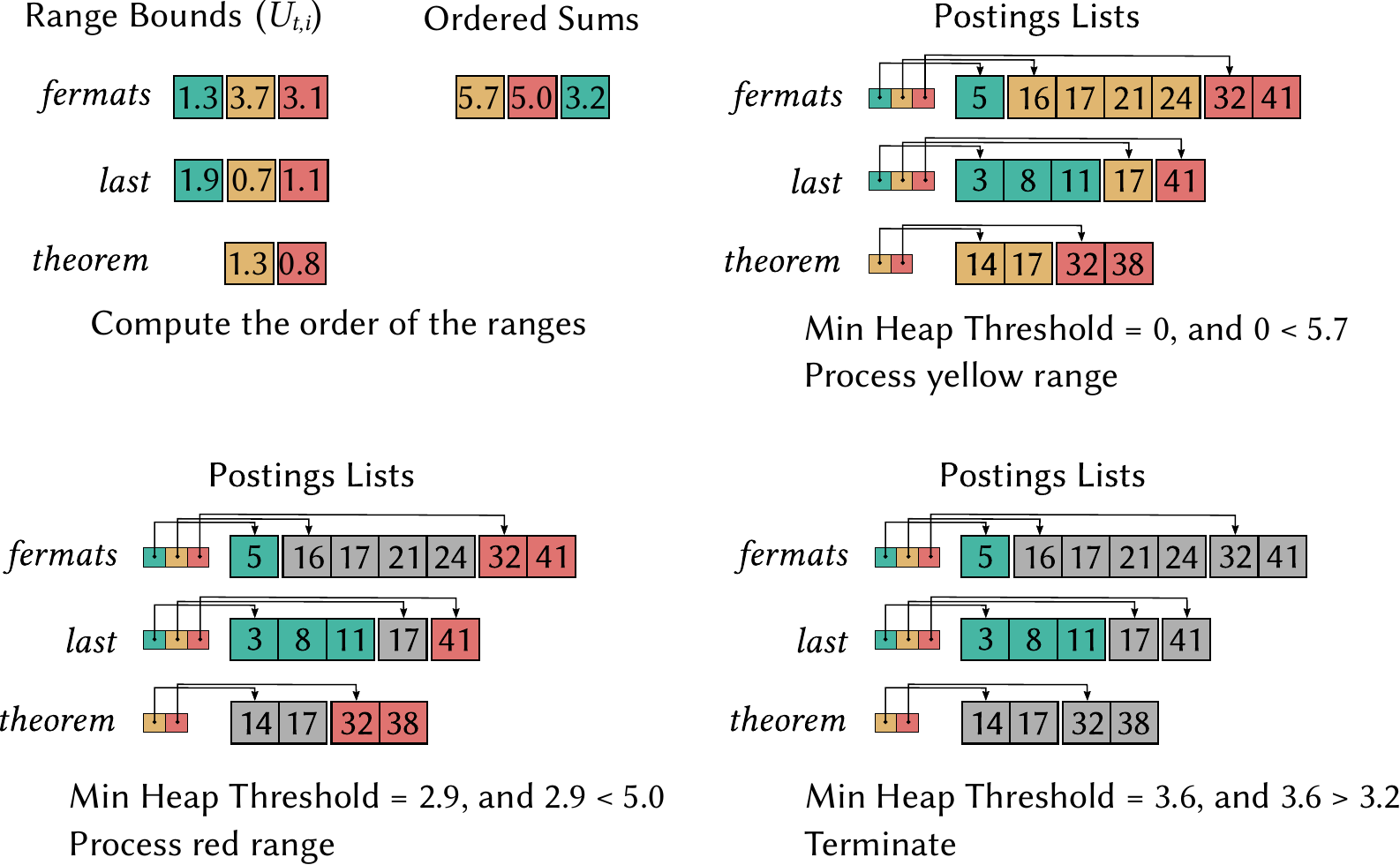}
  \mycaption{Safe early termination for the query {\emph{``fermats
  last theorem''}}.
  To define an ordering over the ranges the query terms' range upper
  bound scores are added and sorted from highest-to-lowest.
  Then, before each range is processed, the current heap threshold
  $\theta$ is checked to determine whether any documents within the
  next range could enter the heap.
  In this example, range~1 (in cyan) does not need to be visited,
  as the min-heap threshold is greater than the upper-bound of the
  range ($3.6 > 3.2$) by the time that range is considered.
\label{fig-fermatquery}}
\end{figure*}

\myparagraph{Early Termination: Safe Ranking}

As well as being motivated by simplicity and fast computation, the
range score bounds can be used to provide rank-safe early termination
(similar to the tier-pruning technique of \citet{dm+17-irj}): given a
query $q$, a current heap score threshold of $\theta$, and a range
$i$, then $i$ (and all subsequent ranges in the decreasing ordering)
can be bypassed if
\[
\sum_{t \in q}{\Termmaxr} \leq \theta \,.
\]
Once this condition becomes true, no document in the $i$\,th range
can make it in to the top-$k$ set for the query, and no document in
any other unprocessed range can either.
All remaining ranges can thus be bypassed.
If this condition is triggered, the result set that is generated is
assured of being {\emph{safe}}, and identical to what would emerge
from processing all of the ranges.
Figure~\ref{fig-example}(c) illustrates such a situation, with the
combined range upper bound score in range~1 less than the heap
bound~$\theta$.

Figure~\ref{fig-fermatquery} builds on Figure~\ref{fig-architecture},
and provides a concrete example of what was illustrated in
Figure~\ref{fig-example}(c).
Each of the three query terms contributes to the set of range upper
bounds, with range~2 (upper bound score $5.7$) being chosen to be
processed first.
Once those documents have been scored, range~3 has the second highest
document score bound, and is processed next.
Once that is done, range~1 can be bypassed, because its highest
possible document of $3.2$ is less than the heap entry threshold
after ranges~2 and~3 have been processed.

\myparagraph{Early Termination: Anytime Ranking}

In some cases, unsafe early termination may be desired.
This is achieved by ending the processing of ranges before reaching
the crossover point between range upper bounds and heap entry score.
The fact that the ranges are processed in order of decreasing score
upper bounds means that the ones processed early should have the best
chance of providing documents into the final top-$k$, and hence
provide the best ``early answers'' if query execution must be halted
and an answer returned.
Section~\ref{sec-experiments} measures the extent to which retrieval
effectiveness is compromised by stopping early.
The critical question that must then be resolved is that of deciding
when, and on what basis, to terminate execution.
Section~\ref{sec-anytime} discusses possible heuristics.

\myparagraph{Improved Pruning With Local Range Bounds}

Dynamic pruning algorithms employ term upper bound scores $U_t$ to
compute limits on each document's maximum score.
In the case of the recent {\bmw} and {\vbmw} methods, those
whole-of-collection bounds are augmented by localized bounds that
apply to blocks of postings pointers, so that the estimation is more
precise, but at the cost of additional index space.

The range upper bound scores required by our proposal can also be
used in that secondary role, and (compared to {\bmw} and {\vbmw})
have the advantage of applying over document ranges rather than over
index postings blocks.
Thus, pivot selection inside each range uses the range-based upper
bounds in lieu of the global bounds, providing more accurate pivot
selection.

\myparagraph{Implementation Issues}

Obvious questions that arise with this proposal include:
\begin{itemize}
\item
	{\newtext{Additional index space:}} the worst-case space cost
	of storing the range upper bounds in the index {\newtext{is
	$\mathcal{O}(r \cdot v)$}}
	\marker{3.13} (there could be $r$
	additional values for each term $t$ {\newtext{in the
	vocabulary $v$}}).
	In practice most terms will only occur
	in a small number of ranges{\newtext{, but even so, the cost
	of storing the bounds might be
	an issue}}.

\item
	{\newtext{Additional time to access postings:}} whether the
	time savings that result from more effective pruning are
	eroded because of non-contiguous index processing.

\item
	{\newtext{Range selection quality:}} the usefulness of the range
	ordering induced by the {\boundsum} heuristic, compared to
	more sophisticated range ordering techniques from selective
	search{\newtext{, might become a factor}}.

\item
	{\newtext{Competitiveness:}}
	how it performs relative to impact-ordered indexes and
	{\saat}-based processing techniques for anytime ranking.

\item
	{\newtext{Meeting an SLA}}:
	how to guide performance so that effectiveness is maximized
	within the constraints dictated by an SLA.

\end{itemize}
{\marker{3.11}}
The next three sections provide experimental results that consider
these five aspects of performance.
 
\section{Experimental Setup}
\label{sec-setup}

{\newtext{This section describes the structure of the main
experiments.
Section~\ref{sec-experiments} then gives results;
Section~\ref{sec-anytime} describes experiments relative to an SLA;
and then Section~\ref{sec-discussion} provides further context in
regard to parallelism and practical throughput.}}
{\marker{3.11}} 

\subsection{Hardware and Software}

All experiments were performed entirely in-memory on a Linux machine
with two 3.50 GHz Intel Xeon Gold 6144 CPUs and 512 GiB of RAM.
Timings were measured as the mean value of three runs.
Document collections were indexed using the
{\sf{Anserini}}~\cite{yfl18-jdiq} system with Porter stemming and
stopping enabled.
Those indexes were then converted to {\pisa}~\cite{pisa19-osirrc} and
{\jassvtwo}~\cite{tc19-spe} indexes using the common index file
format~\cite{lm+20-sigir}, so as to carry out fair comparisons.
All remaining experiments were conducted with the {\pisa} and
{\jassvtwo} search systems.

\subsection{Datasets}

\begin{table}[t]
\centering
\mycaption{
  Details of the two test collections employed.
\label{tbl-collection}}
\begin{tabular}{l S[table-format=8.0] S[table-format=9.0] S[table-format=11.0]}
\toprule
{Corpus} & {Documents} & {Unique Terms} & {Postings} \\
\midrule
{\govtwo}
& 25172934
  & 64675233
   & 5324020638
  \\
{\clueweb}
& 50220189
  & 127466287
    & 14794442493
  \\
\bottomrule
\end{tabular}

 \end{table}

Two public collections are used, {\govtwo} and {\clueweb}, summarized
in Table~\ref{tbl-collection}.
Both have also had {\emph{document clusters}} generated for them
{\citep{dx+16-cikm}}, of which the most competitive are the
{\sf{QKLD-QInit}}
ones, with {\govtwo} and {\clueweb} containing $199$ and $123$ ranges,
respectively.\footnote{\url{http://boston.lti.cs.cmu.edu/appendices/CIKM2016-Dai/}}
A query log was generated by sampling $5{,}000$ random queries from
the TREC {\emph{Million Query Track}} queries {\cite{ac+07-trec, aa+08-trec, cp+09-trec}},
biased so that there are $1{,}000$ queries of each length from $1$ to
$4$ terms, and a further $1{,}000$ queries containing $5$ or more
terms.
The full set of $60{,}000$ queries 
is also used in Section~\ref{subsec-feedback}.

\subsection{System Configurations}

Document ranking is via the {\sf{BM25}} {\cite{rz09fntir}} model with
parameters $k_1=0.4$ and $b=0.9$ {\citep{tjc12-osir}}.
Although many variations of {\sf{BM25}} exist {\cite{tpb14-adcs,
kv+20-ecir}}, {\pisa} and {\jass} use similar formulations and hence
exhibit similar effectiveness {\cite{lm+20-sigir}}.

A document-ordered inverted index is used in {\pisa}, containing
plain {\emph{tf}} values, with all scoring computations taking place
at query time.
These indexes are compressed in fixed-sized blocks of $128$ elements
using the {\simdbp} technique~\cite{lb15-spe}, which gives a good
space/time trade-off~\cite{mss19-ecir}.
Partial blocks shorter than $128$ elements are encoded with
{\emph{binary interpolative coding}} {\cite{ms00-irj}}.
We also tested the {\emph{Partitioned Elias-Fano}} technique {\cite{ov14-sigir}}, 
noting similar trends but with a smaller space occupancy and slower retrieval.
We employ all of {\maxscore}, {\wand}, {\bmw}, and {\vbmw}; the
latter two using fixed (or variable)
sized blocks containing an average of $40$ postings.

In {\jass} impacts are pre-computed and stored quantized using 8-bits
for {\govtwo} and 9-bits for {\clueweb} {\cite{ct+13-cikm}}; with
scoring summing the pre-computed values in an accumulator table.
The impact-ordered postings segments are compressed with
the SIMD accelerated {\emph{Group Elias Gamma}}
{\gegs} {\cite{tl18-adcs}}, which judiciously employs {\sf{VByte}} encoding for 
short lists if it results in better compression than the Elias Gamma code.

\section{Preliminary Experiments}
\label{sec-experiments}

Before examining the performance of the anytime {\daat} arrangement
under strict latency constraints (Section~\ref{sec-anytime}), we
report a number of scoping experiments to establish baseline
relativities, and to gain insights into the underlying performance of
default {\daat} processing, range-based {\daat} processing, and
impact-ordered {\saat} processing.

\subsection{Index Space Consumption}

\begin{table}[t]
\centering
\mycaption{
Space consumption for the three index types across two collections,
in GiB.
For the {\default} document-ordered index, the total storage cost
includes the listwise and variable blockwise upper-bound scores.
Similarly, the {\clustered} document-ordered index cost includes the
listwise, blockwise, and rangewise upper-bound scores, as well as
the additional range mapping data.
For the impact-ordered {\jass} index, only the size of the postings
lists are measured.
{\newtext{Space overheads with respect to the {\default}
document-ordered index for each collection and ordering are shown in
parentheses.}}
{\marker{2.3}}
\label{tbl-space}}
\newcommand{\tabent}[1]{\makebox[12mm][c]{#1}}
\begin{tabular}{l c c c c}
\toprule
  \multirow{2}{*}{Index Type} &  \multicolumn{2}{c}{\govtwo} & \multicolumn{2}{c}{\clueweb} \\
  \cmidrule(lr){2-3}\cmidrule(lr){4-5}
  & {\sf{Random}} & {\sf{Reordered}} & {\sf{Random}} & {\sf{Reordered}}\\
  \midrule
  {\sf{Default}}   & 12.2~\tiny{($1.00\times$)} &\D8.1~\tiny{($1.00\times$)}  & 30.5~\tiny{($1.00\times$)} & 26.1~\tiny{($1.00\times$)} \\
  {\sf{Clustered}} & 15.5~\tiny{($1.27\times$)} &\D9.6~\tiny{($1.19\times$)}  & 35.6~\tiny{($1.17\times$)} & 29.1~\tiny{($1.11\times$)} \\
  {\jass}          & 17.1~\tiny{($1.40\times$)} & 13.5~\tiny{($1.67\times$)} & 46.3~\tiny{($1.52\times$)} & 39.9~\tiny{($1.53\times$)} \\
\bottomrule
\end{tabular}
 \end{table}

Our first experiment quantifies the space overhead incurred by the
cluster skipping inverted index with respect to the default
document-ordered and impact-ordered indexes.
Table~\ref{tbl-space} reports the space consumed by each index type
for {\govtwo} and {\clueweb} indexes, using both {\sf{Random}} and
{\sf{Reordered}} document identifiers, noting that reordering is done
on a global basis for the {\default} index, and on a per-range basis
for the {\clustered} index.
These results confirm the benefits of reordering for document-ordered
indexes {\cite{dk+16-kdd, mm+19-ecir, mss19-ecir}}, but also show the
benefits to impact-ordered indexes, with improvements arising for
both collections and all three index formats.

To understand the specific costs of each index component, the
per-component sizes are as follows.
For {\clueweb}, the baseline globally ordered single range index
requires {\gb{20.8}} for all postings information, and a further
{\gb{5.3}} for the {\wand}/{\bmw} score bounds.
The new range-partitioned index requires {\gb{20.7}} for all postings
information, indicating that topical range-partitioning and local
reordering within clusters has no impact on space usage compared to
global reordering.
In addition to the {\wand}/{\bmw} score bounds, {\gb{3.1}} is
required to store the range upper bounds used by {\boundsum}.
The globally reordered
{\jass} index requires {\gb{39.9}}, substantially more than the
document ordered indexes.
Similar relativities were observed for the {\govtwo} indexes.

\subsection{Effect of Reordering on Score-at-a-Time Retrieval}
\label{sec-jassorder}

{\newtext{I}}t is well established that document
reordering provides substantial benefits to document ordered
indexes.
{\newtext{To determine the effect of document reordering on the
latency of {\sc{SaaT}} retrieval, we employ two indexes: one with the
document identifiers assigned randomly ({\sf{Random}}); and one with
the identifiers assigned according to a global {\bp} ordering
{\cite{dk+16-kdd, mm+19-ecir}}.
We also employ two instances of {\jass}: the first, {\jassex},
exhaustively processes all candidate postings before termination;
whereas the second, {\jassag}, processes a fixed number of postings
($\rho$) before termination.
We use a single value of $\rho$, setting it to $10\%$ of the total
number of documents in the collection {\citep{lt15-ictir}}.
Table~\ref{tbl-jass-order} shows the results.
Reordering document identifiers in the impact-ordered index
arrangement results in substantial speedups (between $1.62\times$ and
$1.76\times$) across the different latency percentiles.
However, {\saat} index traversal is known to exhibit a high correlation
between the number of postings processed and the elapsed latency
{\cite{cclmt17wsdm, lt15-ictir, msc17-adcs}}.
Furthermore, the gains associated with reordering are not diminishing
with the total volume of postings processed, as similar speedups are
observed for both {\jassex} and {\jassag}.
\marker{2.3}}}
So why does reordering save so much time?
{\newtext{I}}nstrumenting the {\jass}
system {\newtext{revealed}} that
reordering the index leads to the high-impact {\newtext{postings for the
terms associated with each query tending to arise in a narrower}}
range of document numbers than otherwise.
Since the {\jass} accumulator table is stored as a two-dimensional
vector {\cite{jto10-adcs}}, localization results in fewer row
initializations (reducing the number of {\tt{memset}} operations),
and substantially fewer cache misses during updates.
Based on these findings, we use the reordered indexes in our
remaining experiments.

\begin{table}[t]
\centering
  \mycaption{The effect of document reordering with {\jass} and
  {\clueweb} ($k=10$), using ``exhaustive'' and ``aggressive''
  processing, reporting median, $95$\,th, and $99$\,th percentile
  latencies in milliseconds per query, {\newtext{and also noting the speedup
  accruing as a result of index reordering.}}
  {\marker{2.4}}
\label{tbl-jass-order}}
\newcommand{\tabent}[1]{\makebox[12mm][c]{#1}}
\begin{tabular}{l S S c S S c}
\toprule
  \multirow{2}{*}{Algorithm} &  \multicolumn{3}{c}{\jassex} & \multicolumn{3}{c}{\jassag, $\rho=10\%$} \\
  \cmidrule(lr){2-4}\cmidrule(lr){5-7}
  & \multicolumn{1}{c}{{\random}} & \multicolumn{1}{c}{\sf{Reordered}} & \multicolumn{1}{c}{\sf{Speedup}}
  & \multicolumn{1}{c}{{\random}} & \multicolumn{1}{c}{\sf{Reordered}} & \multicolumn{1}{c}{\sf{Speedup}}\\
  \midrule
  $P_{50}$ & 106.652 & 60.454  & $1.76\times$ & 65.859 & 37.954 & $1.73\times$ \\
  $P_{95}$ & 345.345 & 201.729 & $1.71\times$ & 77.971 & 47.686 & $1.64\times$ \\
  $P_{99}$ & 489.944 & 296.063 & $1.65\times$ & 82.636 & 50.872 & $1.62\times$ \\
  \bottomrule 
\end{tabular}

 \end{table}

\subsection{Rank-Safe Query Processing}

\begin{figure}[t]
  \centering
  \includegraphics[width=0.70\textwidth]{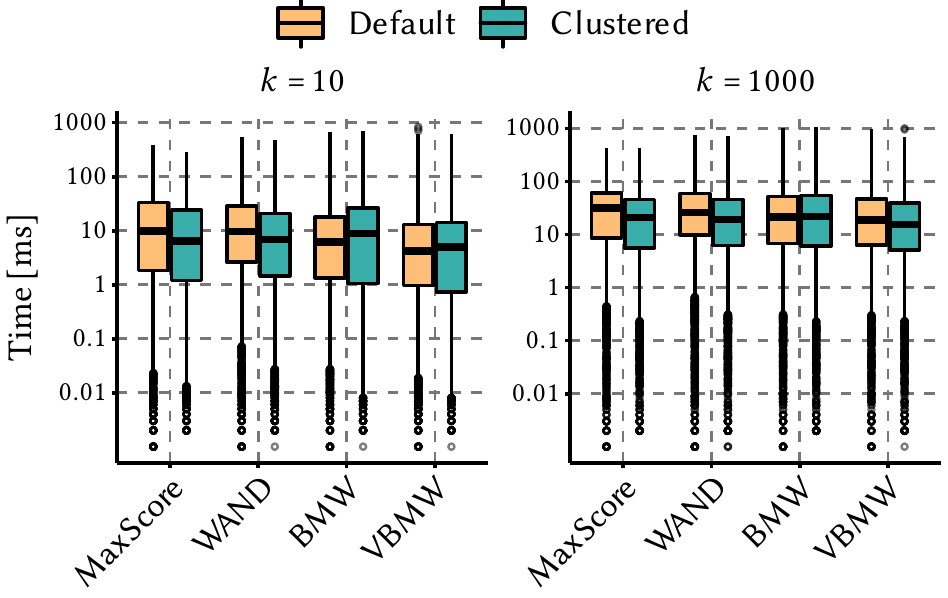}
  \mycaption{Query processing latency with $k=10$ and $k=1000$ for {\daat}
  index traversal algorithms using both the {\default} index with ``standard''
  {\daat} traversal, and the {\clustered} index with range-based traversal.
  \label{fig-safe}
  }
\end{figure}

To gain insight into the performance of the new range-based processing
compared to default {\daat} traversal, we explore rank-safe
processing over the {\clueweb} collection with both $k = 10$ and $k =
1000$.
In this experiment, we use the {\boundsum} technique to determine the
order in which to visit each cluster, and clusters are visited
until the results are guaranteed to be rank-safe.
Figure~\ref{fig-safe} shows the results.
Note that the {\clustered} index arrangement is actually
{\emph{faster}} than the default {\daat} traversal for {\maxscore}
and {\wand} processing.
There are three main reasons for this: firstly, the range-based
algorithms are able to visit high-impact ranges early on in the index
traversal, making the heap threshold rapidly climb towards its final
value (allowing more documents to be skipped); secondly, the query
processing algorithms can exploit the rangewise upper-bounds during
pivot selection, allowing fewer documents to be examined due to more
accurate upper-bound estimates; and thirdly, the range-based
algorithms can exploit safe early termination, allowing entire ranges
to be pruned from the computation.
On the other hand, these three effects have a lesser effect on the
block-based {\bmw} and {\vbmw} techniques, as the local blockwise
upper-bounds provide an even better level of accuracy than the
rangewise upper-bounds.

\subsection{Comparing Range Selection Mechanisms}

The next series of experiments aims to extend the previous analysis
to include {\emph{unsafe}} processing, where efficiency and
effectiveness can be traded off with one another.
In particular, we are interested in evaluating the performance of the
{\boundsum} range ordering technique against a number of baselines to
determine how accurately ranges can be prioritized.

\myparagraph{Range Ordering Baselines}

As a baseline for effectiveness, we compute an {\oracle} range
ordering for each query, applying a geometric weighting to the
documents in the full ranking, and summing over the documents in each
range.
Given a $k$ element {\emph{gold-standard}} ranking $\langle d_1, d_2,
\dots, d_k\rangle$, and a function ${\text{Range}}(d_i)$ which
returns the cluster $c$ containing document $d_i$, then the weight
for each cluster $c_j$ is computed as:
\begin{equation}
\text{Weight}(c_j) = (1-\phi) \cdot \sum^k_{i=1}{x_{i,j} \cdot \phi^{i-1}},
\end{equation}
where $x_{i,j}$ is an indicator variable such that
\begin{equation}
x_{i,j} = 
  \begin{cases}
    1, & \text{if}\ \text{Range}(d_i) = j\\
    0, & \text{otherwise.}
  \end{cases}
\end{equation}
The {\oracle} ranking thus processes ranges in decreasing order of
aggregate rank-biased precision (RBP) weight {\citep{mz08-tois}}.
In the experiments reported shortly a persistence parameter of $\phi
= 0.99$ is used, together with score-safe rankings computed to a
depth of $k = 10{,}000$.
As a second strong baseline, we also measured the effectiveness of
the {\ltrr} range orderings developed by {\citet{dkc17-sigir}} which
employ feature-based learning-to-rank resource selection, and comment
on their practicality below.

\myparagraph{Effectiveness versus Ranges Processed}

\begin{table}[t]
  \centering
  \mycaption{Effectiveness, $k=1000$, of different range orderings
  ({\govtwo}, topics $701$---$850$, and {\vbmw}) with three different
  metrics across various numbers of ranges processed.
  The ``{\oracle}'' ordering is derived from the per-range
  contributions assigned by an {RBP}-like weight distribution applied
  to the full answer ranking.
\label{tbl-better-ra}}
\newcommand{\tabent}[1]{\makebox[9.5mm][c]{#1}}
\begin{tabular}{c c c c c c c c c c c c}
\toprule
  \multirow{2}{*}{Ranges}
  	& \multicolumn{3}{c}{RBP ($\phi = 0.8$)}
		&& \multicolumn{3}{c}{AP@1000}
			&& \multicolumn{3}{c}{RBO ($\phi = 0.99$)}
\\
\cmidrule(lr){2-4}\cmidrule(lr){6-8}\cmidrule(lr){10-12}
& \tabent{\bndsum} & \tabent{\ltrr} & \tabent{\oracle}
  && \tabent{\bndsum} & \tabent{\ltrr} & \tabent{\oracle}
  && \tabent{\bndsum} & \tabent{\ltrr} & \tabent{\oracle}\\
\midrule
  $1$    & 0.412 & 0.541 & 0.559
		&& 0.135 & 0.204 & 0.214
			&& 0.348 & 0.460 & 0.510\\
  $5$    & 0.555 & 0.588 & 0.584 
  		&& 0.252 & 0.283 & 0.288 
			&& 0.698 & 0.770 & 0.861\\
  $10$   & 0.559 & 0.594 & 0.594 
  		&& 0.276 & 0.298 & 0.301 
			&& 0.820 & 0.868 & 0.947\\
  $20$   & 0.588 & 0.594 & 0.594 
  		&& 0.295 & 0.303 & 0.305 
			&& 0.923 & 0.932 & 0.999 \\
  $50$   & 0.595 & 0.594 & 0.594 
  		&& 0.305 & 0.307 & 0.306 
			&& 0.989 & 0.986 & 1.000\\
  All & 0.594 & 0.594 & 0.594 
  		&& 0.306 & 0.306 & 0.306 
			&& 1.000 & 1.000 & 1.000\\
  \bottomrule
\end{tabular}

 \end{table} 

To determine how the number of ranges processed affects the
effectiveness of the different range ordering techniques, we run an
experiment over the {\govtwo} collection with TREC topics
$701$---$850$.
Effectiveness is measured using the {\govtwo} qrels and two metrics:
a shallow instantiation of RBP ($\phi = 0.8$) {\cite{mz08-tois}}; and
the much deeper AP@1000.
We also computed a rank-biased overlap (RBO) score
{\cite{wmz10-tois}} with $\phi=0.99$ relative to a full evaluation,
as a way of measuring similarity of rankings without using qrels.

Table~\ref{tbl-better-ra} confirms that effectiveness rises steadily
as document ranges are processed.
The {\oracle} ordering provides the steepest early growth, but once
around $20$ ranges are processed, its advantage is slender.
Note the strong relationship between RBO (computation of which does
not require qrels) and the shallow and deep effectiveness metrics
(which do).
In the absence of appropriate qrels, RBO is used as a surrogate for
effectiveness in the experiments using {\clueweb} that are presented
below.
{\marker{1.5}}

\myparagraph{Latency versus Ranges Processed}

\begin{figure}[t]
  \centering
  \includegraphics[width=0.60\textwidth]{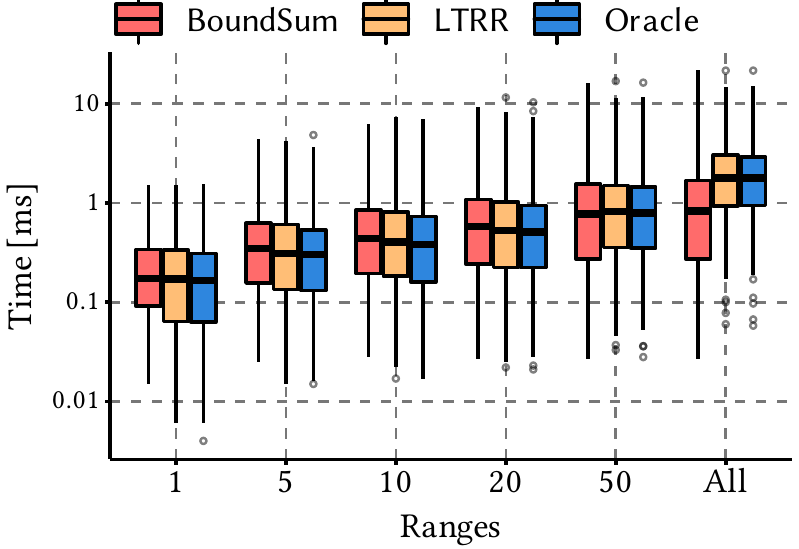}
  \mycaption{Efficiency (msec per query) at
  $k=10$ of different range ordering techniques
  ({\govtwo}, topics $701$---$850$, and {\vbmw}).
\label{fig-better-ra}}
\end{figure} 

To illustrate the relationship between the number of ranges processed
and query latency, Figure~\ref{fig-better-ra} compares the
{\boundsum}, {\ltrr}, and {\oracle} range orderings, plotting query
execution time as a function of the number of ranges processed
({\govtwo} and {\vbmw}, $k=10$).
The {\ltrr} and {\oracle} measurements assume that the range ordering
is available free of cost, whereas the {\boundsum} range ordering
computation is included in the running time; that difference, and the
{\ltrr} and {\oracle} methods' slightly better identification of
fertile ranges, is why they are marginally faster for small numbers
of ranges.
When ``All'' of the ranges are processed, the {\boundsum} approach
benefits from its knowledge of range upper bound scores, and is
usually able to terminate early, an option not possible for the other
two, for which only the range ordering is assumed.

\myparagraph{Efficiency versus Effectiveness}

\begin{figure}[t]
  \includegraphics[width=0.7\columnwidth]{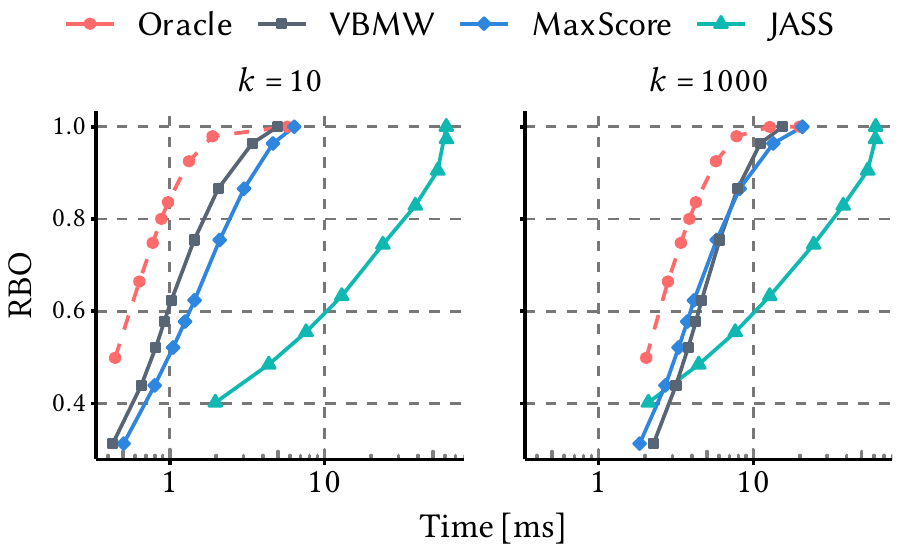}
  \mycaption{Median latency (milliseconds) and RBO ($\phi=0.99)$ for
  {\clueweb}, for two values of $k$.
  The orange dotted lines reflect {\oracle} range orderings, and the
  grey and blue lines the new {\boundsum} technique, all three measured at
  a fixed number of $n$ ranges, with
  $n \in \{1,2,3,4,5,10,20,50,100,123\}$.
  {\newtext{For {\jass}, the plotted points represent $\rho =
  \{0.2,0.5,1,2,5,10,20,50,100\}$ percent of the number
  of documents in the collection.
  Note the log scale on the $x$-axis.}}
  {\marker{1.2}}
  \label{fig-cw09b-sweep}}
\end{figure}

Switching to the larger {\clueweb} collection, and seeking to get a
better understanding of the relativities between the different
techniques, Figure~\ref{fig-cw09b-sweep} plots the trade-off between
effectiveness and query latency that can be achieved by varying the
number of ranges processed {\marker{2.6}}.
{\newtext{We include {\jass} as a further reference point for anytime
querying, with each point corresponding to processing a fixed number
of postings, setting $\rho$ as a fraction of the number of documents
in the collection, varying over the range from $0.2$\% to $100$\%.
The latter might still not guarantee safe evaluation for queries
composed of many common terms, but is a plausible upper bound in
practice and yields an average RBO value greater than $0.999$.}}
{\marker{1.2}} As more ranges are processed, effectiveness increases,
but at the cost of decreased throughput.
The dashed lines show the {\oracle} ordering of the document ranges,
again without any cost attributed to computing the ordering; at the
other side of both graphs, {\jass} is {\newtext{substantially slower
than the {\daat} mechanisms, except for extremely low-latency
contexts where $k$ is large, as {\saat} retrieval is not sensitive to
$k$ {\cite{cclmt17wsdm}.}}}
\marker{1.4} Note that the best choice of the within-range traversal
algorithm varies with $k$: {\vbmw} is fastest for $k=10$, and
{\maxscore} is fastest for $k=1000$.
These results are in line with a recent study showing the superiority
of {\maxscore} for large values of $k$ {\cite{mss19-ecir}}.
Because of that difference, the {\oracle} lines in
Figure~\ref{fig-cw09b-sweep} similarly vary: {\vbmw} is used to
process the {\oracle} range ordering when $k = 10$ and {\maxscore}
when $k = 1000${\newtext{, with the goal of illustrating the best
that could possibly be achieved using the new range-sensitive
approach}}.

\myparagraph{``Enhanced'' Resource Selection?}

Selecting ranges in the clustered index is analogous to selecting
which shards to process in a selective search system, which is why we
also measured {\ltrr}.
However the results shown in Table~\ref{tbl-better-ra} and
Figures~\ref{fig-better-ra} and \ref{fig-cw09b-sweep} present a
rose-tinted view of {\ltrr} and {\oracle} performance, because we
have assumed for both that the range ordering is available at zero
cost.
If even $1$~ms -- an optimistic estimate for the {\ltrr} process,
which involves dozens of parameters per term per document range, and
a learned function that combines them -- is required to compute the
ordering, then the {\boundsum} mechanism provides the superior
trade-off.

\section{Practical Anytime Ranking}
\label{sec-anytime}

Section~\ref{sec-experiments} established that document-ordered
indexes are able to prioritize ``fertile'' document ranges.
Now we consider how to employ that approach to ensure adherence to
strict latency-based SLAs.
Our main experiment targets two situations, $P_{99}\leq50$~ms, and
$P_{99}\leq25$~ms, chosen to suit likely first-stage retrieval
expectation of a single node in a large distributed search system
that utilizes multiple cascading rankers.
But to show the versatility of the new approach, we also consider a
third ``stretch'' target of $P_{99}\leq10$~ms.

\subsection{Decision-Making}

The first step in a practical anytime ranking implementation is to
determine the decision-making process.

\myparagraph{Termination Points}

A key aspect of anytime query processing is knowing {\emph{when}} to
terminate.
Checking a termination condition may itself incur a time penalty (for
example, a system call to compute elapsed time, or pausing a loop to
check a posting count) and needs to be invoked sparingly, balancing
fidelity against overheads.
In the anytime {\jass} system, the termination condition is checked
between segments {\cite{lt15-ictir}}, even if they have the same
impact, which provides a suitable fidelity for accurate termination
without adding significant overheads.
We choose to check termination conditions prior to the start of each
document range, with the granularity then depending on the time taken
to process the query against a single range.

\myparagraph{Online Latency Monitoring}

Previous anytime ranking implementations employed linear predictive
models based on counts of postings to determine how much work can be
done by a particular system relative to some fixed time budget
{\cite{lt15-ictir, msc17-adcs}}.
These techniques work well when predicting {\saat} latency, because
the innermost loop of {\saat} retrieval simply involves decoding a
segment, and processing all postings within it.
In {\daat} scoring, however, the processing time depends on a wide
range of factors, including term co-occurrence and relative term
importance, where the high-score documents arise in the postings, and
even the ranking function used {\cite{pcm13-adcs, mto12-sigir}}.
Furthermore, predictive models may need dozens of features to achieve
accurate predictions {\cite{sk15-wsdm, mto12-sigir, tm20acmtois}},
and may not be adaptive to external factors such as system load
{\cite{db13-cikm}}.

Instead of predicting latency, we first simply {\emph{monitor}} it as
each query is being processed.
We use {\texttt{std::chrono::steady\_clock}} to measure the elapsed
time for each document range, and use those measurements to make a
sequence of ``go/no-go'' decisions.
Microbenchmarking revealed that these calls take around
$2.5$~microseconds each, meaning that the total overhead when
processing $20$ ranges (see Table~\ref{tbl-better-ra}) is around
$0.05$~milliseconds.

\myparagraph{Termination Policies}

Deciding when to terminate query processing is important to achieve
maximal effectiveness without violating the latency SLA.
While obvious policies such as termination after processing $n$
ranges ({\fixed}-$n$) are available, they are not sensitive to
important factors such as query length, term density, or even system
load.
Online monitoring enables a number of improved strategies, by
explicitly capturing budgets in real time.

Suppose that $B$ is the SLA latency budget, that $t_i$ is the
measured time spent processing the first $i$ document ranges, and
that a decision must be taken on whether to {\sf{Terminate}}, or
{\sf{Continue}} to process range $i+1$.

One obvious policy, {\greedy}, is defined as follows:
\begin{equation}
  {\greedy}(t_i, B) = 
  \begin{cases}
    {\textsf{Continue}}, & \text{if}\ t_i < B\\
    {\textsf{Terminate}}, & \text{otherwise.}
  \end{cases}
\end{equation}
Essentially, {\greedy} risks violating the SLA by one range's worth of 
processing, and relies on the range upper bound-based pruning to meet the SLA
overall. 

A second option, {\cautious},  is defined as follows:
\begin{equation}
  {\cautious}(t_i, B, \tmax) = 
  \begin{cases}
    {\textsf{Continue}}, & \text{if}\ t_i + \tmax < B\\
    {\textsf{Terminate}}, & \text{otherwise,}
  \end{cases}
\end{equation}
where $\tmax$ represents some absolute upper bound on the cost of
processing any query against any range, which can be set based on the
relative granularity of the range processing latency.
While {\cautious} will always remain within the SLA, effectiveness
might be lost because of its pessimism -- time deemed to be available
might not be used.
In our experiments, we fix $\tmax=5$~ms, but note the possibility
that $\tmax$ might be tuned based on other factors, such as the
underlying hardware, or the maximum range size, and so on.

The problem with both {\greedy} and {\cautious} is that they rely on
fixed time values, and do not adapt to the specific features of the
query being processed.
On the other hand, monitoring the latency during query processing
provides rich information about the relative difficulty of processing
a given query, allowing better estimates to be made.
Thus, we propose a third policy, {\judicious}:
\begin{equation}
  {\judicious}(t_i, B, \alpha) = 
  \begin{cases}
    {\textsf{Continue}}, & \text{if}\ t_i + \alpha ({t_i}/{i}) < B\\
    {\textsf{Terminate}}, & \text{otherwise,}
  \end{cases}
\end{equation}
where $\alpha\ge1$ is an adjustment that allows different SLA and
load scenarios to be accommodated, with the next range processed only
if the available time remaining to the query is at least $\alpha$
times the mean per-range processing time observed so far.
For example, when $\alpha=1$, the mean range time so far for this
query is used as the predicted time of the next range.

Taking $\alpha=1$ is optimistic, and might result in SLA violations
on as many as half of the queries in highly constrained operational
environments, an unacceptably high fraction.
Indeed, what is actually required is not just an accurate estimate of
the mean time to process each range, but also an estimate of the
variance of those per-range times.
To allow for volatility about the mean, values $\alpha>1$ can be
used.
In the next subsection we compare performance using the fixed value
$\alpha=1$; Section~\ref{subsec-alpha} compares $\alpha=1$ and
$\alpha=2$; and finally Section~\ref{subsec-feedback} describes a
process for establishing $\alpha$ on the fly, using an adaptive
feedback mechanism that is reactive to the localized query workload.

\subsection{SLA Compliance}

\begin{table}[t]
  \centering
    \mycaption{Anytime processing on {\clueweb} ($5{,}000$ queries,
    {\vbmw}, $k=10$ queries) and two different SLA time limits,
    measuring effectiveness using average RBO ($\phi=0.8$).
    The $P_{50}$, $P_{95}$, and $P_{99}$ columns list the $50$\,th
    (median), $95$\,th, and $99$\,th percentile query execution times
    across the query set.
    The ``Miss'' and ``\% Miss'' columns represent the number of
    and percentage of queries which violate
    the SLA, and the ``Mean'' and ``Max'' columns report the mean and
    maximum amounts by which those misses exceed the SLA.
    Note that throughout this section, {\jass} makes use of a
    reordered index, discussed in Section~\ref{sec-jassorder}.
    \label{tbl-latency-final}}
  \newcommand{\tabent}[1]{\makebox[7mm][c]{#1}}
\renewcommand{\tabcolsep}{0.7em}
\sisetup{
group-separator = {,},
round-mode=places,
round-precision=1,
table-format=2.1,
round-minimum=0.1,
}
\begin{tabular}{
l
S S S[table-format=2.1] S[table-format=4.0] S[table-format=<2.1] S[table-format=<2.1] S[table-format=3.1] S[table-format=1.3,round-precision=3] }
\toprule
\multirow{2}{*}{System}
	& \multicolumn{8}{c}{SLA: $P_{99} \leq 50$~ms}
\\
\cmidrule{2-9}
& {\tabent{$P_{50}$}}
	& {\tabent{$P_{95}$}}
		& {\tabent{$P_{99}$}}
			& {Miss}
				& {\% Miss}
					& {Mean}
						& {Max}
							& {RBO}
\\
\midrule
{\sf{Baseline}} {\vbmw}
& 4.151 
	& 50.063 
		& \violatex 116.3 
			& \violatex 244 
				& \violatex 5.0
					& 53.621 
						& 786.582 
							& 1.000
\\
{{\fixed}\sf{-123}}
& 4.982 
	& 45.732 
		&\violatex 96.2  
			& \violatex 210
				& \violatex 4.3
					& 40.267 
						& 568.882 
							& 1.000
\\
{\etvbmw}
 & 4.345
 & 49.921
 & \violatex 50.0
 & \violatex 241
 & \violatex 5.0
 & 0.039
 & 0.096
 & 0.974
 \\
\midrule
{\sf{JASS-5}}
& 38.55 
	& 48.820 
		& \violatex 54.0  
			& \violatex 190
				& \violatex 3.9
					& 3.020  
						& 13.562  
							& 0.841
\\
{\sf{JASS-2.5}}
& 23.79 
	& 29.414 
		& \withinx 31.5
			& \withinx 0
				& \withinx 0.00000
					& {---}
						& {---}
							& 0.751
\\
{{\fixed}\sf{-20}}
& 2.056 
	& 18.416 
		& \withinx 39.5  
			& \withinx 28
				& \withinx 0.0560
					& 21.317 
						& 132.531 
							& 0.900
\\
{{\fixed}\sf{-10}}
& 1.440 
	& 12.434 
		& \withinx 25.3  
			& \withinx 8
				& \withinx 0.00160
					& 14.416 
						& 34.680  
							& 0.812
\\
\midrule
{\greedy}     
& 5.645 
	& 47.827 
		& \violatex 50.6 
			& \violatex 217
				& \violatex 4.5
					& 0.620  
						& 9.035   
							& 0.991
\\
{\cautious}   
& 5.647 
	& 45.072 
		& \withinx 45.9  
			& \withinx 1
				& \withinx 0.00020
					& 0.301  
						& 0.301   
							& 0.990
\\
{\judicious}, $\alpha=1$  
& 4.953 
	& 44.796 
		& \withinx 49.6  
			& \withinx 12
				& \withinx 0.00240
					& 0.894  
						& 3.453   
							& 0.990
\\
\bottomrule
\\
\toprule
\multirow{2}{*}{System} & \multicolumn{8}{c}{SLA: $P_{99} \leq 25$~ms}\\
\cmidrule{2-9}
& {\tabent{$P_{50}$}}
	& {\tabent{$P_{95}$}}
		& {\tabent{$P_{99}$}}
			& {Miss}
				& {\% Miss}
					& {Mean}
						& {Max}
							& {RBO}
\\
\midrule
{\sf{Baseline}} {\vbmw}
& 4.151 
	& 50.063 
		& \violatex 116.3
			& \violatex 614
				& \violatex 12.7 
					& 37.237 
						& 811.582 
							& 1.000
\\
{{\fixed}\sf{-123}}
& 4.982 
	& 45.732 
		& \violatex 96.2 
			& \violatex 621
				& \violatex 12.8 
					& 28.160 
						& 593.882 
							& 1.000
\\
{\etvbmw}
 & 4.266
 & 25.039
 & \violatex 25.1
 & \violatex 613
 & \violatex 12.7
 & 0.037
 & 0.135
 & 0.928
 \\
\midrule
{\sf{JASS-5}}
& 38.55 
	& 48.820 
		& \violatex 54.0 
			& \violatex 3427
				& \violatex 70.6 
					& 15.906 
						& 38.562  
							& 0.841
\\
{\sf{JASS-2.5}}
& 23.79 
	& 29.414 
		& \violatex 31.5
			& \violatex 1893
				& \violatex 39.0 
					& 2.333  
						& 9.296   
							& 0.751
\\
{{\fixed}\sf{-20}}
& 2.056 
	& 18.416 
		& \violatex 39.5 
			& \violatex 132
				& \violatex 2.7  
					& 16.090 
						& 157.531 
							& 0.900
\\
{{\fixed}\sf{-10}}
& 1.440 
	& 12.434 
		& \violatex 25.3 
			& \violatex 51
				& \violatex 1.0  
					& 13.508 
						& 59.680  
							& 0.812
\\
\midrule
{\greedy}
& 5.642 
	& 25.278 
		& \violatex 26.6 
			& \violatex 665
				& \violatex 13.7 
					& 0.478  
						& 9.227   
							& 0.978
\\
{\cautious}
& 5.633 
	& 20.370 
		& \withinx 21.9  
			& \withinx 10
				& \withinx 0.0020
					& 1.967  
						& 4.654   
							& 0.970
\\
{\judicious}, $\alpha=1$
& 4.941 
	& 24.767 
		& \withinx 24.9  
			& \withinx 34
				& \withinx 0.0680
					& 0.839  
						& 2.952   
							& 0.976
\\
\bottomrule  
\end{tabular}
 \end{table}

Our next experiment determines the trade-offs between the various
anytime processing regimes and termination policies.
Table~\ref{tbl-latency-final} shows how these various alternatives
perform against two latency SLAs and a 99\% conformance requirement.
Blue numbers in the table indicate SLA compliance, and red values
indicate violations.

At the top of the table, the {\sf{Baseline}} {\vbmw} which uses a
standard document-ordered inverted index does not provide any SLA
guarantees, and nor does {\fixed}{\sf{-123}} (named because there are
$123$ ranges in total in {\clueweb}), which processes all available
ranges in the order specified by {\boundsum}.
Both of these provide rank safe processing.
{\newtext{We also experimented with an {\emph{early-terminating}}
{\vbmw} approach ({\etvbmw}) which executes the {\sf{Baseline}}
{\vbmw} technique, but checks a timer every $10{,}000$ iterations of
the main {\vbmw} control loop, and terminates if the SLA is exceeded.
This {\etvbmw} method misses the target SLAs by small amounts
(primarily because it operates on an {\greedy} termination policy);
more importantly, it suffers from non-trivial effectiveness loss,
because high-scoring documents for the query might occur near the end
of the collection and never get scored.}}
{\marker{2.2}}

The middle portion of the table considers methods that terminate
after a fixed amount of work, and include {\jass} ($\rho=5$ and $2.5$
million postings per query), {\newtext{and {\fixed} (processing the
top $20$ or $10$ ranges in the order specified by {\boundsum}).}}
{\marker{2.7}}
They work moderately well at $50$~ms, but are non-compliant against
the more aggressive $25$~ms target, and further tuning of the bounds
would be required.
The two {\jass} versions and {\fixed}{\sf{-10}} also have RBO scores
that are notably lower {\newtext{than the RBO achieved by the
{\fixed}{\sf{-20}} approach.}}
{\marker{2.7}} 

The last section in each of the two main blocks of the table shows
the advantage of the {\judicious} monitoring approach, here with
$\alpha=1$.
Unsurprisingly, {\greedy} misses both the $50$~ms and $25$~ms SLAs,
as it only terminates once the allocated time budget is exceeded.
On the other hand, both {\cautious} and {\judicious} ($\alpha=1$) are
compliant, with (as expected) {\cautious} further away from the
actual SLA bounds, and hence with (at $25$~ms) inferior RBO scores.
Similar results arise for $k = 1000$.

\subsection{Incorporating Variance for Strict SLAs}
\label{subsec-alpha}

\begin{figure}[t]
  \centering
  \includegraphics[width=0.48\textwidth]{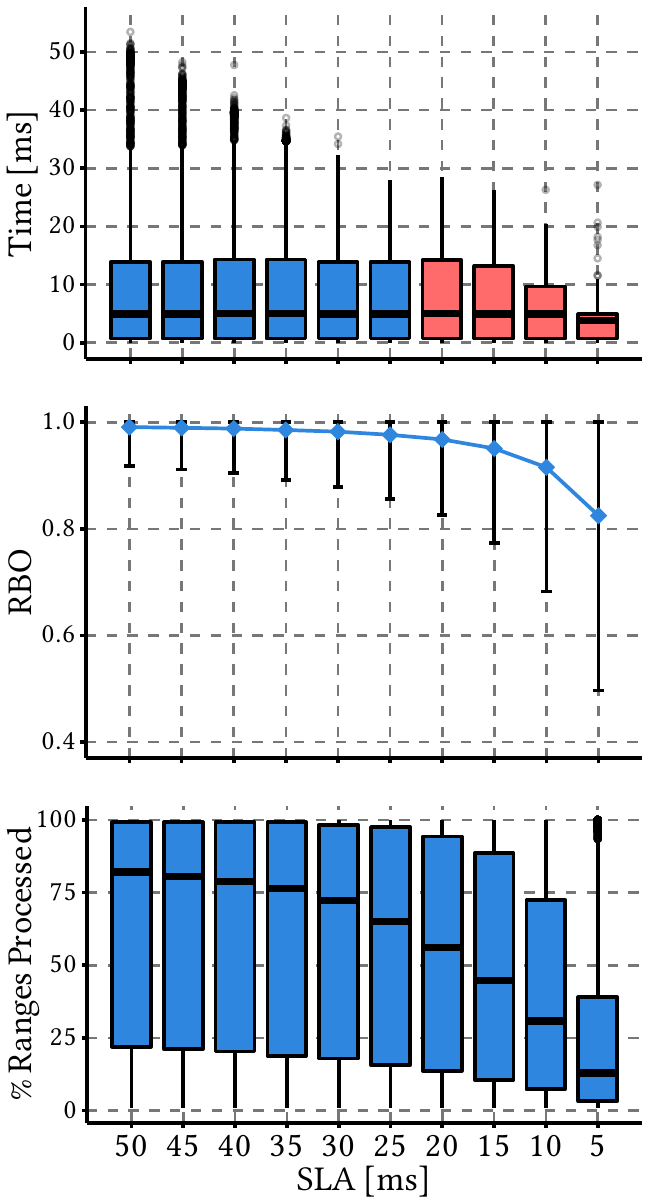}
  \includegraphics[width=0.48\textwidth]{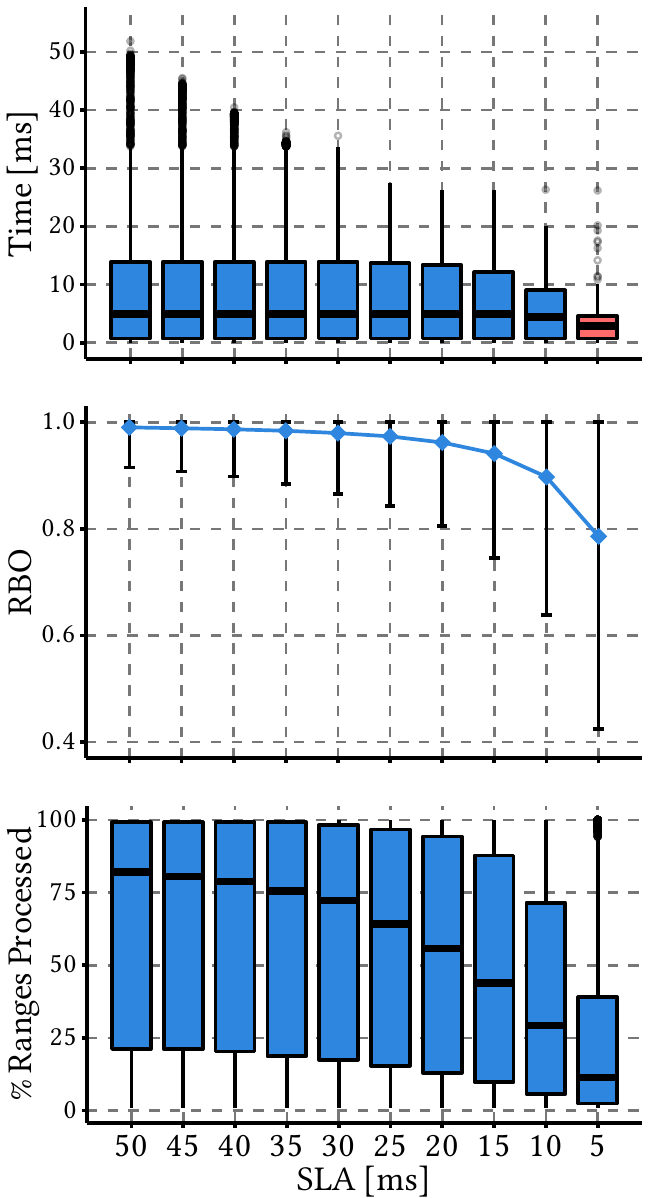}
  \mycaption{Execution time and RBO profiles for {\vbmw} index
  traversal over {\clueweb} with $k = 10$ for different $P_{99}$
  SLAs, using the {\judicious} policy with $\alpha = 1.0$ (all left-side plots) and
  $\alpha = 2.0$ (all right-side plots).
  In the latency plots in the top row, red boxes denote a violation
  of the SLA time limit shown on the $x$-axis.
  In the two RBO plots in the middle row, the error bars correspond
  to the standard deviation, with the upper value truncated at $1.0$.
  By shifting to $\alpha=2.0$, allowance is made for volatility in the
  estimation process used to predict the execution time of the next
  range, allowing stricter SLAs to be met in practice.
  The two panes in the bottom row show the fraction of available
  ranges (of a total of at most $123$ for {\clueweb}) that are
  processed before the query is denied further execution resources.
  \label{fig-alpha}
  }
\end{figure}

To demonstrate the role of $\alpha$, Figure~\ref{fig-alpha} plots
latency and RBO values for {\vbmw} processing over {\clueweb} as
increasingly strict SLA requirements are introduced.
While $\alpha = 1$ is able to meet a $P_{99}$ SLA of $25$~ms, it
fails for any stricter SLA.
A more conservative predictor (setting $\alpha=2$) allows a sequence
of increasingly tight SLAs to be met, down to (but not including)
$P_{99} \leq 5$~ms.
The cost is that $\alpha=2$ also results in a decrease in the
relative RBO score, as less overall work is completed prior to
termination.
When the SLA is set at $5$~ms only around $10$\% of the document
ranges associated with each query get processed on average, shown in
the bottom row of Figure~\ref{fig-alpha}.
As a consequence, the similarity computation is increasingly
disrupted.
More critically, the estimation process itself breaks down at this
severe SLA -- more than $1$\% of the queries exceed the $5$~ms limit
after just one range has been processed.
That is, $5$~ms $P_{99}$ SLA compliance would only be possible if the
collection were reconfigured into more ranges -- for example, by
splitting each of the current ranges into two halves.

\begin{figure}[t]
  \centering
  \includegraphics[width=0.65\textwidth]{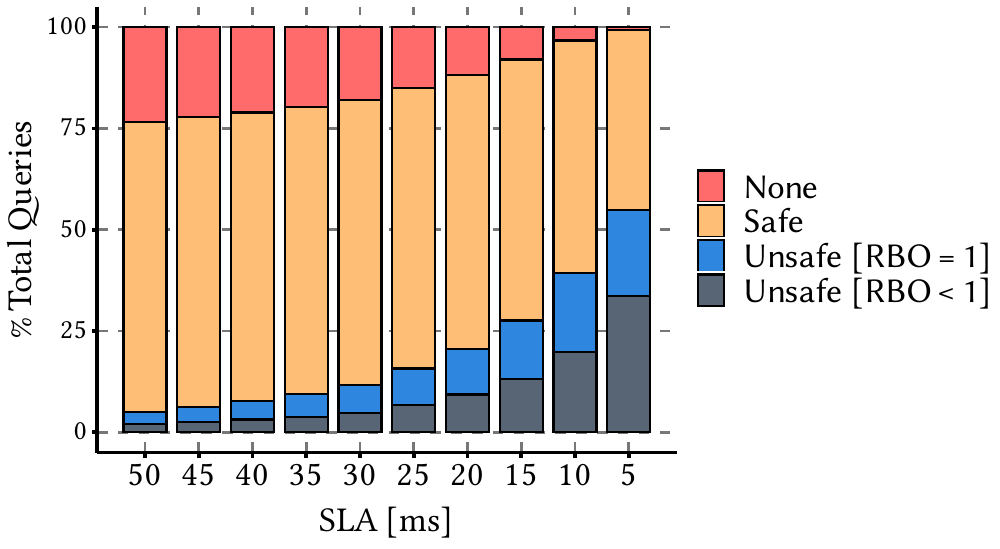}
\mycaption{The evolving split between complete query execution,
  score-safe early termination (as a result of the range bounds and
  range processing order), and unsafe termination, as the SLA bound
  is made increasingly onerous, plotted for {\clueweb}, {\judicious}
  processing, and $\alpha=1.0$.
  \label{fig-profile}}
\end{figure}

Figure~\ref{fig-profile} illustrates these issues from another
perspective.
When the SLA is $50$~ms, many queries run through all of the
associated ranges, or are terminated safely.
Both of these groups generate RBO scores of $1.0$.
But as the SLA is made more acute, an increasing fraction of queries
must be terminated unsafely, and each time that occurs, retrieval
effectiveness is eroded.
Note, however, that even when the rank safety of the results cannot
be guaranteed (when the processing is terminated early based on the
given termination policy), the results may already be the same as the
exhaustive top-$k$ results.
At an SLA of $5$~ms more than half of all queries require unsafe
termination to be applied; moreover, as already noted, the SLA does
not get met.

\subsection{Choosing $\alpha$}
\label{subsec-feedback}

Having introduced $\alpha$ as part of the {\judicious} mechanism, the
obvious question is that of selecting a suitable value.
Too small, and we risk violating the SLA.
Too large, and we risk ending processing prematurely, with SERP
quality then put at risk -- as is visible in Figure~\ref{fig-alpha}.
Moreover, if the SLA is defined in terms of (say) $P_{99}$, then it
is not just permissible to have $1$\% of the queries exceed the
stipulated bound, but it is actively {\emph{desirable}} to do so,
rather than over-achieving and having all queries complete within the
limit.
That is, the SLA can be viewed as providing a target as well as
imposing a limit.
To that end, we introduce a further mechanism, defined as:
\begin{equation}
  {\adaptive}(t_i, B, \alpha, \beta) = 
  {\judicious}(t_i, B, \alpha)
\end{equation}
in terms of how the ``go/no-go'' decisions get made, but with an
{\emph{adjustment step}} performed after each query completes, to
provide feedback into the value of $\alpha$, defined via
\begin{equation}
  \alpha \leftarrow
  \begin{cases}
    \alpha \times \beta & \text{if}\ t_i > B \, ,\\
    \alpha \times (1/\beta)^{Q} & \text{otherwise}\, ,
  \end{cases}
\end{equation}
where, as before, $t_i$ is the time taken by that query to process
the first $i$ ranges, where $Q$ is the SLA tolerance level (for
example, $Q=0.01$ if the bound is based on $P_{99}$ tail latency),
and where $\beta>1$ determines the aggressiveness of the reactive
feedback process.
In this approach, each time a query exceeds the SLA time limit,
$\alpha$ increases by the multiplicative factor $\beta$, as a signal
to be more conservative with upcoming queries.
And each time a query ends within the SLA time limit, $\alpha$ is
(very slightly) relaxed.
For example, if $\beta=1.5$, then each ``within-limit'' query reduces
$\alpha$ by a factor of $0.995953558$, and a sequence of $100$
within-limit queries cause $\alpha$ to decrease by a factor of
$0.995953558^{100}=2/3$.
On the other hand, any ``limit exceeded'' query immediately causes
$\alpha$ to grow by $50$\% -- the logic being that, if at all
possible, each too-slow query should be followed by approximately
$100$ within-limit ones before another too-slow one can be tolerated.

\begin{figure}[t]
  \centering
  \includegraphics[width=0.7\textwidth]{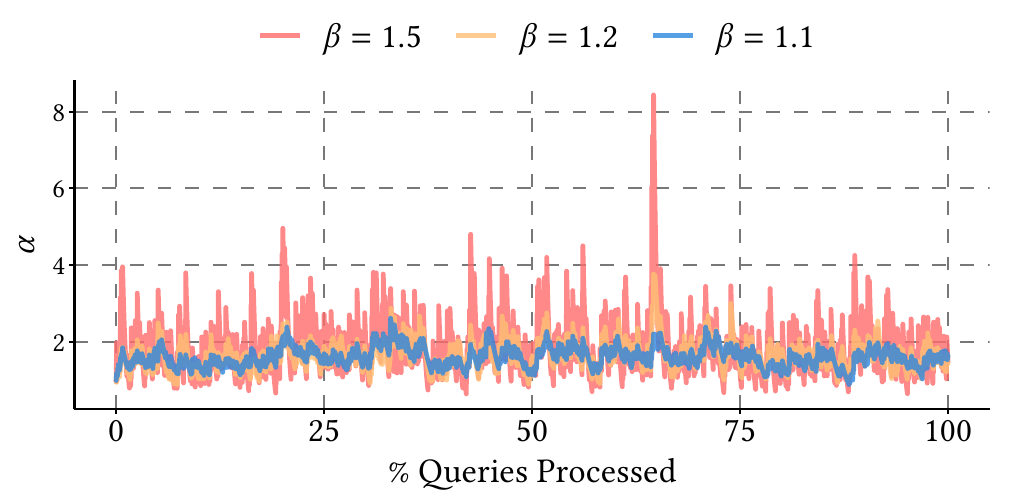}
  \includegraphics[width=0.7\textwidth]{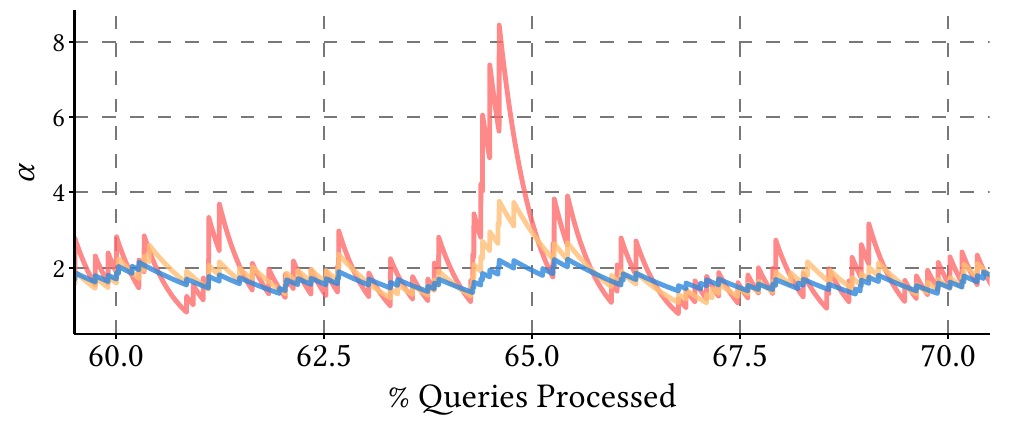}
    \mycaption{Reactive anytime processing on {\clueweb} ({\vbmw},
    $k=10$) and an SLA defined by $P_{99}\leq 10$~ms, applied across
    all $60{,}000$ Million Query Track queries.
    The bottom pane is an expansion of a small part of the top pane,
    and shows the sawtooth nature of the adjustments that take place
    as $\alpha$ reacts to the query stream.
    \label{fig-flame}}
\end{figure}

Figure~\ref{fig-flame} shows the effect of the {\adaptive} strategy,
tracing $\alpha$ as queries are processed, using three different
values of $\beta$, the full sequence of $60{,}000$ Million Query
Track queries, and an aggressive SLA target of $P_{99}\leq 10$~ms.
When $\beta=1.5$, changes to $\alpha$ are relatively rapid, and the
values used tend to ``hunt'' somewhat.
The smaller values of $\beta$ still allow rapid adjustments to be
made to $\alpha$ when local load spikes are encountered in the query
stream, but give more consistent behavior overall.
In this configuration the trend $\alpha$ value is between one and
two, but it sometimes also rises above two as part of the feedback
process.

\begin{table}[t]
  \centering
    \mycaption{Anytime processing on {\clueweb} ({\vbmw}, $k=10$, and
    $60{,}000$ queries) and an SLA time limit of $10$~ms.
The columns have the same meaning as in
    Table~\ref{tbl-latency-final}, but the values cannot be compared
    because of the different query sequence employed.
    \label{tbl-vary-alpha}}
  \newcommand{\tabent}[1]{\makebox[7mm][c]{#1}}
\renewcommand{\tabcolsep}{0.7em}
\sisetup{
group-separator = {,},
round-mode=places,
round-precision=1,
table-format=2.1,
}
\begin{tabular}{
l
S[table-format=1.1] S[table-format=1.1] S[table-format=2.1] S[table-format=4.0] S[table-format=1.1] S[table-format=1.1] S[table-format=2.1] S[table-format=1.3,
                                                        round-precision=3]
}
\toprule
\multirow{2}{*}{System}
	& \multicolumn{8}{c}{SLA: $P_{99} \leq 10$~ms}
\\
\cmidrule{2-9}
& {\tabent{$P_{50}$}}
	& {\tabent{$P_{95}$}}
		& {\tabent{$P_{99}$}}
			& {Miss}
				& {\% Miss}
					& {Mean}
						& {Max}
							& {RBO}
\\
\midrule
{\judicious}, $\alpha = 1.0$ 
& 3.800
	& 9.949
		& \violatex 10.5
			& \violatex 1495
				& \violatex 2.5
					& 1.013
						& 25.696
							& 0.934
\\
{\judicious}, $\alpha = 2.0$ 
& 4.021
	& 9.821
		& \withinx 9.9
			& \withinx 363
				& \withinx 0.6
					& 2.113
						& 25.639
							& 0.925
\\
{\adaptive}, $\beta = 1.5$
& 3.650
	& 9.863
		& \withinx 10.0
			& \withinx 595
				& \withinx 1.0 
					& 1.596
						& 25.512
							& 0.928
\\
{\adaptive}, $\beta = 1.2$ 
& 3.733
	& 9.870
		& \withinx 10.0
			& \withinx 596
				& \withinx 1.0 
					& 1.660
						& 25.436
							& 0.931
\\
{\adaptive}, $\beta = 1.1$
& 3.761
	& 9.869
		& \withinx 10.0
			& \withinx 599
				& \withinx 1.0
					& 1.635
						& 28.232
							& 0.931
\\
\bottomrule
\end{tabular}
 \end{table}

Table~\ref{tbl-vary-alpha} mimics Table~\ref{tbl-latency-final}, but
retains the strict SLA of $P_{99}\leq 10$~ms used in
Figure~\ref{fig-flame}, and now focuses on the {\adaptive} strategy.
On this longer query sequence the three {\adaptive} methods (with
$\beta=1.5$, $1.2$, and $1.1$) all provide SLA compliance, with
marginally higher RBO scores than the previous
(Figure~\ref{fig-alpha}) ``guess'' of $\alpha=2$.
The volatility in $\alpha$ when $\beta=1.5$ affects the RBO scores,
and among the three {\adaptive} approaches, the more stable $\alpha$
is, the greater the average RBO score that results.
A key benefit of the new {\adaptive} approach is thus that it
provides automatic load shedding in high pressure situations, as a
consequence of the way it dynamically adapts to query latency on a
trailing query-by-query basis.

\subsection{Further Comment}

Finally in this section, note that these experiments are for a single
collection and two query streams derived from the same initial
source, and practitioners might need to further tune the available
latency/effectiveness trade-off space for their specific search
deployment.
More sophisticated reactive estimation techniques that take into
account other factors such as the number of query terms could also be
readily added to this framework.
For example, an array of $\alpha$ variables might be maintained, so
that each reacts to -- and hence learns a suitable value for --
queries of different term counts, or containing terms with different
collection frequencies.

Nor have we sought yet to exploit the flexibility that might be
gained by deciding which subset of $1$\% of the queries might be
``sacrificed'' while still meeting the SLA, and whether, once those
queries go past the time limit, further processing effort is
allocated to them.
For example, if an SLA is specified via two clauses, requiring (say)
$P_{99}\leq 10$~ms and $P_{100}\leq 20$~ms, then it might be possible
to estimate in advance which $1$\% of the queries would usefully
benefit from the extra $10$~ms, and include such guidance into the
go/no-go decision made after each range is processed.
That is, the {\judicious} and {\adaptive} implementations reported
here terminate each query immediately after it goes past the $P_{99}$ limit
point, but that might not be the best overall strategy.
We plan to examine these options in future work.
 
\section{Discussion and Validation}
\label{sec-discussion}

It was observed in Section~\ref{sec-introduction} that retrieval
systems storing large amounts of data are, of necessity, partitioned
across a cluster of machines, with each machine responsible for a
subset of the collection; and that partitioning has the beneficial
side-effect of reducing latency, because when the data is partitioned
much of the work associated with each query can be carried out in
parallel.
It was also observed that where the query load is higher than can be
managed by one instance of the collection (whether partitioned or
not), the collection would be replicated, with each replicate able to
respond to queries independently of the others.
This latter point implies that operating costs are minimized if each
cluster is as fully query-loaded as is consistent with the SLA; while
the former observation means that operating costs are minimized if
each cluster has as few machines as are consistent with meeting the
SLA.

\subsection{Scale and Workload}

The total workload expected of a system is (broadly) proportional to
the product of the size of the collection, perhaps measured in TiB,
and the number of queries to be processed.
On the other hand, the resources available to deliver that total work
are the number of processors, summed across all replicates,
multiplied by the time spent processing the query stream, perhaps
measured in seconds.
That is, the appropriate measure for distributed IR system throughput
is ``(TiB $\times$ queries) / (machines $\times$
seconds)''~{\citep{mz04wise}}.

Given that relationship, it is clear that (within reasonable limits)
any given SLA can be met by adding more processors to each cluster,
and making each partition smaller.
But the total hardware cost likely increases if that is done, because
each machine is carrying less data than it could, and hence measured
performance in ``(TiB $\times$ queries) / (machines $\times$ seconds)''
is likely to decrease.
That is, adding hardware to an efficiently-balanced system is an
``expenditure-based'' way of meeting the additional constraints
introduced by an SLA.
That is why in this paper we have explored algorithmic approaches to
SLA compliance.

\subsection{Partitioned Collections}

Having shown in Sections~\ref{sec-experiments} and~\ref{sec-anytime}
that one machine and one collection can be effectively managed on one
server so as to comply with an aggressive SLA, we still need to also
show that all of the servers in a cluster -- each holding a share of
the total document collection -- can be similarly managed, so that
the cluster as a whole will meet a given SLA.
To do that, further experimentation was undertaken, focusing solely
on partitioning, and noting that replication takes care of
itself.

The standard approach to partitioning index data across a pool of
servers is by {\emph{random assignment}}, whereby each document is
allocated to one of the available servers based on a hash value
derived from the document identifier, or from a pseudo-random value.
The benefits of random allocation are well documented, and include
highly-effective load balancing, preventing individual machines from
becoming bottlenecks.
Each query that arrives at the cluster is to sent all of the servers;
they all compute and return an answer set for it; and then those
answer sets are merged to form a single ranking.
As a thought experiment, suppose that the {\clueweb} collection that
has been used here as the main experimental context is actually a
random $1/P$\,th subset of some even larger resource, call it
{\bigclueweb}, a collection that is $P$ times the size of {\clueweb}.
We would like the machine hosting ``our'' fraction of the larger
collection to remain synchronized with the other $P-1$ machines
hosting the other partitions, so that the SLA applying to the whole
of {\bigclueweb} would be met.

We do not have access to that hypothetical {\bigclueweb}
collection.
But suppose further that a decision was made that {\bigclueweb} was
to be split across $2P$ processors, rather than~$P$.
Each processor would then manage half of the current {\clueweb},
receiving, in effect, a random $50$\% subset of the documents; and
thus should be able to process queries approximately twice as
quickly.
Measurement of the stability (or not) of query execution times on
random halves of {\clueweb} would then provide a basis for inference
in regard to the stability of the single-processor results that were
already presented for {\clueweb}, and hence allow conclusions about
the stability of the hypothetical cluster of $P$ processors that is
taking responsibility for {\bigclueweb}.

\begin{table}[t]
  \centering
    \mycaption{Mean and range of three efficiency measures and one
    effectiveness measure for anytime processing over $10$ different
    random $50$\% subsets of {\clueweb}.
    Processing is conducted with {\vbmw} using the {\judicious}
    policy ($\alpha = 2.0$), with $k=10$, a set of $5{,}000$ queries,
    and five different SLA targets.
    \label{tbl-validate}}
  \newcommand{\tabent}[1]{\makebox[12mm][c]{#1}}
\begin{tabular}{c c c c c c c c c}
  \toprule
  \multirow{2}{*}{$P_{99}$ SLA} & \multicolumn{2}{c}{$P_{50}$} & \multicolumn{2}{c}{$P_{95}$}  & \multicolumn{2}{c}{$P_{99}$}  & \multicolumn{2}{c}{RBO}\\
  \cmidrule(lr){2-3}
  \cmidrule(lr){4-5}
  \cmidrule(lr){6-7}
  \cmidrule(lr){8-9}
  & Mean & Range & Mean & Range & Mean & Range & Mean & Range \\
  \midrule
$25$~ms & 3.8 & 0.12 & 24.3  & 0.04 & 24.6  & 0.03 & 0.967 & 0.003\\
  $20$~ms & 3.8 & 0.12 & 19.5  & 0.04 & 19.7  & 0.03 & 0.954 & 0.003\\
  $15$~ms & 3.8 & 0.14 & 14.6  & 0.03 & 14.8  & 0.02 & 0.916 & 0.047\\ 
  $10$~ms & 3.7 & 0.08 & \D9.8 & 0.02 & \D9.9 & 0.03 & 0.879 & 0.025\\
 \D$5$~ms & 2.6 & 0.05 & \D4.9 & 0.01 & \D5.3 & 0.29 & 0.767 & 0.014\\
 
  \bottomrule
\end{tabular}
 \end{table}

With that purpose in mind, ten random subsets of {\clueweb} were
generated, each consisting of $50\%$ of the documents.
Then, for each of those random subcollections, the experiment
presented in Figure~\ref{fig-alpha} was carried out in its entirety,
using {\vbmw}, the {\judicious} termination policy with $\alpha=2$,
and a range of appropriate latency SLAs limits, measuring in each
case actual query latencies and RBO scores.
Table~\ref{tbl-validate} shows the mean value of each those two
quantities across the set of ten trials, and with a
$\var{max}-\var{min}$ (range) span also reported for each measured
quantity.
As can be seen in the table, across a range of SLA levels the ten
subcollections demonstrate consistently small $\var{max}-\var{min}$
ranges, typically less than $3$\% of the value being measured.
That is, the set of ten random ``half {\clueweb}'' collections all
behave consistently, providing evidence in support of our claim from
Section~\ref{sec-introduction} that it is sufficient to focus on one
non-partitioned collection and one machine, provided that the
collection can be assumed to be a random subset drawn from whatever
larger set of documents makes up the ``full'' collection.

\subsection{Parallel Processing}\marker{1.1}

\begin{figure}[t]
  \centering
  \includegraphics[width=0.49\textwidth]{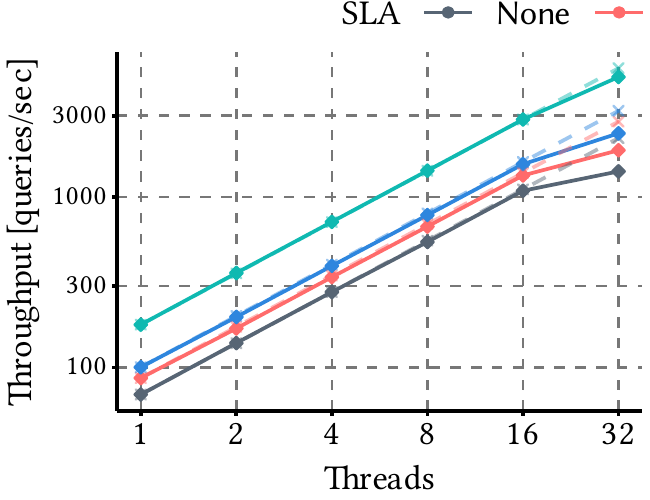}\hfill
  \includegraphics[width=0.49\textwidth]{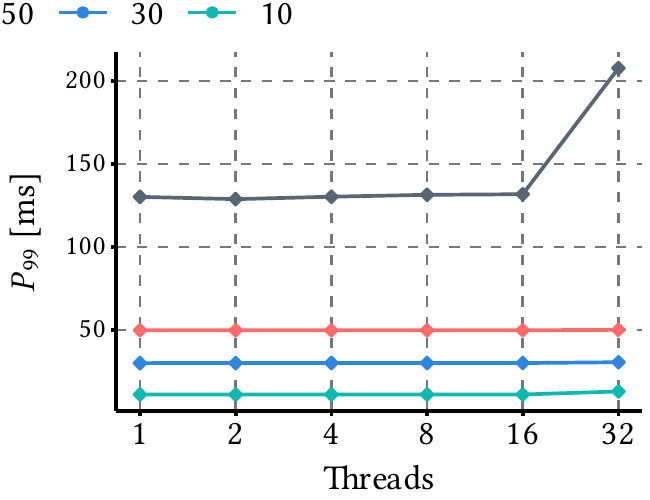}
  \mycaption{\newtext{Throughput (left, total queries per second) and
  $P_{99}$ tail latency (right, milliseconds) for {\vbmw} processing
  of {\clueweb} with $k = 10$ and different $P_{99}$ SLAs, using the
  {\judicious} policy with $\alpha = 1.0$, plotted as a function of
  the number of concurrent processing threads, on a server with $32$
  cores.
  In the left figure, the dashed lines represent {\emph{perfect
  parallelism}}; the solid lines the observed performance.
  {\marker{1.1}}
  \label{fig-parallel}
  }}
\end{figure}

\noindent
{\newtext{We also explored the scalability of our proposed approach,
using parallel execution on a server with $16$ cores and support of up to $32$
concurrent threads sharing a single
memory, to determine how inter-processor contention affects total
query throughput at different SLA levels.
To simulate a uniform query load, each active processing thread was
assigned a randomly permuted copy of the input query log ($5{,}000$
queries, see Section~\ref{sec-setup}), with the number of active
threads controlled as part of the experiment.
Figure~\ref{fig-parallel} shows the net throughput of {\vbmw} with
the {\judicious} termination policy for four SLAs (including no SLA)
on the left, and measured $P_{99}$ tail latency on the right.
Throughput with the new anytime {\daat} approach scales almost
linearly with the number of processing threads, while maintaining
strict pre-stipulated SLAs.
Throughput is sublinear (shown by the dashed lines) only when the
machine is fully loaded ($32$ concurrent queries), because the system
itself preempts the query processing threads so that its own tasks
can be carried out.}}

\subsection{Failure Analysis}

To better understand cases where the clustered anytime processing
fails, we also examined some queries in detail, sampling from across
the range of RBO scores.
Table~\ref{tbl-failure} lists fifteen queries, ordered by decreasing
RBO (the basis on which they were selected), and provides a number of
indicators for each query, in all cases based on two top-$10$
retrieval runs, first an exhaustive computation, and then an
SLA-limited one with a $P_{99}$ target of $25$~ms.
For example, the first query, about poverty in Tucson, had
all of the true top-$10$ documents located in a single document range
(the denominator in column one) and that range was processed by the
SLA-limited {\boundsum} implementation (the numerator in column one);
there were a total of $123$ ranges processed by {\judicious} (column
two); the deepest position in the {\boundsum} ordering of any of the
(in this case, just one) answer-bearing ranges was in position $3$
(column three); and the average depth of the answer-bearing clusters
was also $3$ (column four).
For this query the dynamic pruning process was not able to provide
safe early termination even after the single answer-bearing range had
been processed; fortunately, each per-range execution was fast enough
that all ranges could be processed within the $25$~ms SLA bound.
The second and third queries in the first group were also processed
in a score-safe manner, with the the third query terminated (either
by the {\boundsum}-based pruning being sure that safety had been
attained, or by the time limit being reached and the safety being by
chance) after only half the ranges had been processed.

\begin{table}[t]
  \centering
    \mycaption{Sample queries at five different points in the RBO
    spectrum relative to exhaustive evaluation, assuming anytime
    processing over {\clueweb}.
    Processing is conducted with {\vbmw} using the {\judicious}
    policy ($\alpha=1.0$), with $k=10$, and an SLA target of $25$~ms.
    The columns record (left to right) the number of answer-bearing
    ranges processed by the SLA-limited implementation, expressed as
    a fraction of the number identified by exhaustive top-$10$
    retrieval, ``Ans.''; the number of ranges processed by
    {\judicious} for that query, ``Proc.''; the depth of the deepest
    answer-bearing range in the {\boundsum} range ordering,
    ``Dpst.''; and the average depth in the {\boundsum} range
    ordering of the answer-bearing ranges.
    \label{tbl-failure}}
  \newcommand{\localq}[1]{{\small\emph{#1}}}
\begin{tabular}{@{}c c c c c l@{}}
\toprule
  \multicolumn{4}{c}{Document Ranges} && \multirow{2}{*}{Query}\\
  \cmidrule{1-4}
   Ans. & Proc. & Dpst. & Avg.\\
   \midrule
\multicolumn{4}{c}{RBO${}= 1.00$}\\
   \cmidrule{1-4}
1/1 & 123 &\D\D3 & \D3.0 &&
	\localq{what is poverty level tucson az}
\\
2/2 & 110 &\D\D2 & \D1.5 &&
	\localq{return of capital in mutual fund}
\\
3/3 & \D64 &\D13 & \D5.7 &&
	\localq{women's suffrage lesson plans}
\\[0.5ex]
\midrule
  \multicolumn{4}{c}{RBO${}= 0.75$}\\
   \cmidrule{1-4}

5/6 & \D84 &\D85 & 38.5 &&
	\localq{when is irs going to release frozen refunds}
\\
4/5 & \D58 &\D77 & 33.4 &&
	\localq{state of michigan building codes}
\\
2/4 & \D\D8 &\D35 & 13.8 &&
	\localq{social welfare policy public law 100-485 title ii.}
\\[0.5ex]
\midrule
  \multicolumn{4}{c}{RBO${}= 0.50$}\\
   \cmidrule{1-4}

2/6 & \D18 &\D96 & 34.0 &&
	\localq{lost money washington state}
\\
4/7 & \D12 &\D41 & 15.0 & &
	\localq{changing social security number because of marriage}
\\
1/3 & \D\D9  &\D22 & 14.6 &&
	\localq{child protective service laws washington state}
\\[0.5ex]
\midrule
  \multicolumn{4}{c}{RBO${}= 0.25$}\\
   \cmidrule{1-4}

5/7 & \D43 &\D83 & 36.7 &&
	\localq{child support collection rates in ct}
\\
1/8 & \D19 &\D92 & 60.4 &&
	\localq{what can a special education teacher do beside teach}
\\
2/5 & \D\D9 &\D59 & 22.4 &&
	\localq{clear view school day treatment program}
\\[0.5ex]
\midrule
  \multicolumn{4}{c}{RBO${}= 0.00$}\\
   \cmidrule{1-4}

0/4 & \D\D8 & 121 & 65.0 & &
	\localq{most popular names from social security}
\\
0/5 & \D\D5 &\D71 & 41.6 &&
	\localq{family medical leave my company fired me over this}
\\
0/3 & \D\D1 &\D40 & 34.0 &&
	\localq{how does analyzing a process help improve quality
		within an organization}
\\
\bottomrule
\end{tabular}
 \end{table}

Moving down the table, the general trend is for a decreasing number
of ranges to be able to be processed per query within the available
time, and hence for an increasing fraction of the required (for
top-$10$ fidelity) ranges to be missed.
These problematic queries are also characterized by an absence of
topical coherence, with the target set of top-$10$ documents spread
across as many as eight different ranges, and those ranges themselves
being scattered through the {\boundsum} ordering, as noted by the
column labeled ``Avg.''
In extreme cases, slow per-range querying and lack of topical focus
is a combination that can lead to critical failure.
For example, the final query in Table~\ref{tbl-failure} shows this
combination.
By the time the first range has been processed, the {\judicious}
policy terminates processing to adhere to the strict SLA, and none of
the answer-bearing ranges are examined.

There are also cases where the RBO score is low even when a majority
of answer-bearing ranges are visited, such as the first query in the
RBO${}=0.25$ band, about child support in Connecticut.
Four of the seven required ranges (to get the correct top-$10$) were
visited, but the three which were not contained the documents at
ranks $\{1, 2, 3, 5, 7\}$ in the full ranking, hence the low RBO
score.

As a final comment, note that while a low RBO score indicates that
the documents being returned are not the same as those returned by an
exhaustive system, it is entirely possible that the retrieved
documents are still useful to the user in terms of answering the
query.
In future work we will develop ways to carry out effectiveness
experiments, hoping to develop more accurate techniques for
identifying relevant clusters that lead to even better performance;
while noting the delicate balance between shard selection accuracy
and the time spent generating such selections already discussed in
connection with Figure~\ref{fig-cw09b-sweep}.

\section{Conclusion and Future Work}
\label{sec-conclusion}

To keep up with increasingly large volumes of data, search
practitioners require techniques which allow efficiency and
effectiveness to be traded off in a controllable manner.
We have described new methods for carrying out anytime ranking, where
the quality of the ranking increases with the amount of computation
undertaken; and have showed that strict latency budgets can be
adhered to by monitoring query processing on-the-fly, allowing better
effectiveness outcomes under such budgets.
Unlike prior work using impact-ordered indexes and score-at-a-time
traversal, our approach employs document-ordered indexes and
document-at-a-time retrieval, processing the index via
topically-focused document ranges, and selecting a processing order
for each query using a simple heuristic.
Our experimental analysis shows that the range-based approach
outperforms a series of baselines including impact-ordered indexes,
and is capable of meeting relatively strict SLAs over large document
collections.

There are a number of avenues for further investigation.
Firstly, the clustering technique employed for building the index
clusters was drawn from the selective search literature, and combined
with the recursive graph bisection index reordering technique.
However for our purposes document reordering and document clustering
are closely related, and better techniques for assigning documents to
clusters and ordering the documents within each cluster may be worth
investigating.
Secondly, more precise variants of the {\judicious} and {\adaptive}
policies may be possible, perhaps by exploiting the variance of the
range processing times to improve the accuracy of the next-range
latency estimation, or by managing multiple values of~$\alpha$, or by
being strategic in regard to which queries are allowed to exceed the
SLA time limit.
{\newtext{Thirdly, recent work from {\citet{tm20acmtois}} showed that
a small sample of the inverted index, known as an {\emph{index
synopsis}}, can be used to predict query latency.
It might be interesting to explore how index synopses could be
employed for clustered indexes, and whether further benefits could be
realized with respect to the latency monitoring approaches presented
here.}}
{\marker{2.1}}
{\newtext{Finally, we note that the ideas explored in
this work might translate to other query processing paradigms,
including impact-ordered indexing and score-at-a-time retrieval.}
\marker{1.4}} We look forward to tackling these problems in future work.

\myparagraph{Software}
The software that was developed during this work, and additional data and 
tools, are available at {\url{https://github.com/jmmackenzie/anytime-daat}}.

\myparagraph{Acknowledgment}
We thank the reviewers for their detailed comments which helped improve the manuscript.
This work was supported by the Australian Research Council Discovery Project DP200103136.

\renewcommand{\bibsep}{2.5pt}
\bibliographystyle{abbrvnat}

\end{document}